\documentclass{aa}  

\usepackage{graphicx}
\usepackage{txfonts}
\usepackage{soul}
\usepackage{color}
\renewcommand\hl[1]{#1}

\usepackage[pdfpagelabels=false]{hyperref}      
\hypersetup{colorlinks=true,linkcolor=blue,citecolor=blue,filecolor=blue,urlcolor=blue,}
\begin{document}

   \title{Coevolution of intracluster light and brightest cluster galaxies}

   \author{Rebecca J. Mayes, 
          \inst{1}
        Facundo A. G\'omez
        \inst{1}
          \and
          Antonela Monachesi\inst{1}
          }

   \institute{Departamento de Astronomía, Universidad de La Serena, Avda. Raúl Bitrán Nº 1305, La Serena, Chile\\
              \email{rebecca.jane@userena.cl}
             \thanks{}
             }

   \date{Received x; accepted y}

\abstract{Intracluster light (ICL) is a faint stellar component of galaxy groups and clusters bound to the cluster potential, making up a significant fraction of the cluster mass. The ICL formation and evolution is strongly linked to the brightest cluster galaxies (BCGs) of clusters.}{We aim to compare the properties and progenitor galaxies of the ICL and BCGs of clusters and groups at redshift z = 0 and to determine their processes of coevolution.}{We selected 127 clusters and groups in the hydrodynamic IllustrisTNG-100 simulation above a mass of 10\textsuperscript{13}~\(\textup{M}_\odot\). We divided the ICL from the BCG by applying a surface brightness cut at the Holmberg radius of 26.5 mag/arcsec\textsuperscript{-2}, where star particles within this radius are defined as being attached to the BCG. Those beyond outside are defined as being part of the ICL. We  studied the properties and formation history of the ICL and BCG.}{We find the ICL is generally composed of material from stripped or merged intermediate-mass galaxies, with a smaller in situ component, whereas the BCG is composed of more massive merged galaxies and has a higher in situ fraction. The ICL mass fraction increases weakly with cluster mass, declines with concentration, and increases with time since the time of the BCG's most recent major merger. The ICL is bluer and more metal-poor than the BCG, but there is no significant difference in the age of the material. Universally, BCG+ICL systems have negative colour and metallicity gradients. The ICL and BCG share a high fraction of progenitor galaxies, but the most significant progenitor is typically unknown.}{The ICL properties and formation are tied to the formation histories of the host cluster and BCG and, thus, their properties are individual to each system. Although the ICL and BCG are coevolved, they have distinct formation histories and properties.}

    \titlerunning{Coevolution of ICL and BCGs}
    \authorrunning{Rebecca J. Mayes, Facundo A. G\'omez\and Antonela Monachesi\inst{1}}
   \maketitle

\section{Introduction}
Intracluster light (ICL), also known as diffuse light, is a faint stellar component of galaxy groups and clusters that is not bound to any individual galaxy in the cluster. Instead, it is bound to the cluster potential itself and extends for hundreds of kiloparsecs from the cluster centre. The existence of the ICL was first theorised and then later discovered by \citet{Zwicky1937, Zwicky1951} in the Coma cluster. The ICL has since been found across a wide range of different groups, clusters, and redshifts \citep[e.g.][]{montes2022}. Many theoretical and observational studies have been conducted to study the formation of the ICL and its link to the formation and evolution of its host cluster (see \citealt{contini2024} for a recent review). The ICL makes up a significant fraction of cluster mass (between 5 to 50 per cent), which varies between clusters and also has a strong dependence on how the ICL is defined \citep[e.g.][]{Mihos_2015ICL, contini2021b, montes2022}. No clear dependence has been found for cluster mass \citep{montes2022, Ragusa_2023} or redshift \hl{\mbox{\citep{montes2022, Joo_2023}}}.

The formation of the ICL is strongly linked to brightest cluster galaxy (BCG) formation and evolution. BCGs are located close to the minima of the potential well in host clusters \citep{Jones1984}, making their evolution closely linked to that of their host clusters. They have properties that are distinct from those of typical elliptical galaxies (\citealt{contini2024} and references therein) and are thought to grow primarily by mergers rather than in situ star formation \citep[e.g.][]{Aragon1998, Lidman2012, Burke_2013, OlivaAltamirano2014}.

The BCG and the ICL develop similarly, growing by mergers and stellar stripping. Both are also linked to cluster development. However, they can have quite different properties and can be distinguished based on these properties, as the ICL is \hl{younger}, bluer, and metal-poorer than the BCG \citep{Morishita2017, Montes2018, DeMaio2018, Contini2019}. 

There are a few different formation mechanisms that can contribute to the ICL. These include: (1) in situ star formation, which \hl{was initially proposed} by \citet{Puchwein2010} {but has not been found in observations} \citep{Melnick_2012} as there is minimal ongoing star formation in ICL regions. Some recent {simulation} studies have suggested that in situ stars can still make up a significant percentage of the ICL and extend for hundreds of kiloparsecs from the cluster centre \citep{Ahvazi2024}. {The only observational evidence of in situ star formation in the ICL comes from }\hl{\mbox{\citet{Barfety_2022},} who observed in situ star formation in the ICL in a cluster at z=1.7}. Galaxy growth mechanisms can also cause in situ stars to be found in the ICL; for example, major mergers that disrupt the \hl{galaxy} can eject in situ material from the central areas of the BCG \hl{\mbox{\citep{Willman_2004, Conroy_2007}}}; (2) galaxy mergers, where a satellite is fully disrupted within the potential well of the cluster, usually due to a merger with the BCG, which typically takes place in more dense environments  (see def.1; \citealt{Monaco2006, Murante2007,contini2014}); and (3) stellar stripping, which occurs when stars belonging to a satellite galaxy are stripped by tidal forces, but the satellite galaxy remains bound \citep{Rudick2009, Rudick_2011, Montes_2014, Contini_2018,DeMaio2018,Montes2018,contini2024}. The primary formation mechanisms of the ICL \hl{have been found in simulations} to be the stellar stripping and mergers of intermediate and massive galaxies \citep{Monaco2006, Murante2007, Somerville2008,Guo2011,Martel2012,contini2014,Contini_2018, Brown2024}. Ongoing questions remain about the origins and masses of galaxies stripped to form the ICL, as well as how these properties might vary depending on the environment and the different merger histories a cluster may follow.

\hl{Simulations} have established that the main contributors to ICL formation are expected to be high-mass \hl{\mbox{(>10\textsuperscript{10}~\(\textup{M}_\odot\))}} galaxies \citep{Contini2019,contini2024,Chun2023, jeon2025} and the formation can be heavily influenced by a small number of massive progenitors \citep{Cooper20150}. For example, \citet{Harris_2017} simulated a Fornax-like cluster and found >60 per cent of the z = 0 ICL formed from just two massive objects. \citet{Ahvazi2024} studied ICL formation in the TNG50 simulation and found that half the intragroup light (IGL) and ICL across considered systems originated in galaxies with stellar masses between 10\textsuperscript{10} and 10\textsuperscript{11}~\(\textup{M}_\odot\).

Quantifying the contribution of stellar stripping and mergers separately can be challenging, as the definition of when a merger takes place can vary, as discussed in \citet{Contini_2018}. \citet{Murante2007} and \citet{contini2014} find different contributions due to different definitions of ongoing mergers. In semi-analytic models such as \citet{contini2014} a merger is defined as occurring when a satellite cannot be distinguished from the central galaxy, so they have become the same object. However, usually, before the merger, the satellite undergoes tidal stripping, and these stripped stars can be considered as formed from a merger. \citet{Contini_2018} showed that by relaxing the merger definition used in \citet{contini2014} their contributions aligned with \citet{Murante2007}. A secondary mechanism is pre-processing, where the ICL is formed in the group environment and later merges with the host cluster. Up to 30 per cent of the ICL is formed by this mechanism \citep{contini2014}. However, in this scenario, the ICL was formed earlier by methods, such as mergers or stripping, and then accreted, making this a sub-channel, rather than a primary mechanism. 

The ICL has been found in some environments and simulations to build up at low redshifts, with stripping being the dominant formation channel at z~$\approx$~0 \citep[e.g.][]{Montes_2014, Montes2018, demaio2015, DeMaio2018, Golden_Marx_2023,contini2024}.
\hl{Formation at high redshifts is more uncertain, with mergers potentially becoming the more dominant component around \mbox{z$\sim$1}} \citep{Joo_2023, Jimnez_Teja_2023, Brown2024}. However, in general, the properties of the ICL will be strongly linked with the formation history and dynamical state of the cluster, with the ICL serving as a fossil record for cluster dynamics \citep{Jim_nez_Teja_2018,Contreras_Santos_2024}.

Recent observational studies have suggested that the ICL can also be used as a luminous tracer of the underlying dark matter distribution, in large part due to its formation based on dynamical interactions \citep[e.g.][]{Montes2019, Alonso_Asensio_2020,Yoo_2024} and because it is comprised of stars influenced only by the potential well of the dark matter distribution. The ICL is more adept at tracing the dark matter distribution than other proxies, such as X-ray measurements and gas \citep{Montes2019, Yoo_2024}.

One major issue in considering the properties of the ICL is that separating it from the BCG is very difficult, especially in observations. Several different mechanisms can be used to divide the ICL and the BCG, often producing divergent results \citep[e.g.][]{Brough2023}. Isolating the ICL in observations first requires deep and wide observations of extended low-surface-brightness areas, removing all contaminants and masking satellite galaxies, after which the BCG and the ICL must be separated. In simulations, the primary challenge is separating the BCG and the ICL. Different mechanisms for separating the ICL and the BCG include: (1) a simple radius cut, usually of 30 kpc or 100 kpc, where all stellar material outside the cut is assumed to be ICL; (2) a cut based on the surface brightness profile of the BCG, generally set at the Holmberg radius of 26.5 mag/arcsec\textsuperscript{-2} \citep{Rudick_2011}; and (3) {more complex multi-component surface-brightness-profile fitting that models the BCG, the ICL and also allows for a characterisation of the transition zone} \citep[e.g.][]{Janowiecki_2010,iodice2017, kravtsov2018, Zhang2019}. A triple S\'ersic profile fitting is generally used for this method \hl{\mbox{\citep{Spavone_2018, spavone2024}}}. The disadvantage of this method is that it is heavily reliant on the specific functional forms chosen. Machine learning can help to speed up otherwise computationally expensive classifications \citep{Marini_2022}. Increasingly, due to the difficulty of dividing the BCG and the ICL, many studies have considered the two as a single system \citep{kravtsov2018,DeMaio2018,Zhang2019,demaio2020}. Because the two often have different properties, they are usually distinguished based on examining properties such as \hl{surface brightness profiles, }age, colour, and metallicity \citep{Gonzalez2005, Gonzalez_2007, Montes_2014, Montes2018, Montes2021, Montes2022a, Morishita2017,DeMaio2018,Contini2019, SampaioSantos2021, Golden_Marx_2023, goldenmarx2025, Brough2023, Zhang2024, Kimmig_2025, Brough2023}.

An ongoing question is how the ICL and BCG growth are linked and how the ICL and BCG coevolve (or otherwise) depending on formation history and properties. In this paper, we investigate the coevolution of the \hl{ICL} and \hl{BCG} in the IllustrisTNG-100 simulation, with the aim of determining how the properties of the ICL and BCG are linked and dependent on their formation histories and the properties of their host cluster. This paper is organised as follows. Section \ref{section:methods} describes the simulations used and methods of ICL analysis. Section \ref{section:ICLmassfrac} compares ICL mass fraction to cluster properties. Section \ref{section:ICLBCGcomps} discusses the components that form the ICL and BCG, while Sect. \ref{section:compcorr} compares these components to cluster properties. Section \ref{section:ICLBCGprops} presents a discussion of the ICL and BCG metallicities colours and ages. Section \ref{section:ICLprogs} discusses ICL and BCG assembly histories. Section \ref{section:conc} summarises the main conclusions.

\section{Methods}
\label{section:methods}
In this section we give an overview of the IllustrisTNG-100 simulations used to analyse the ICL and BCG and the methods used for selection and analysis. Table~\ref{tab:defs} defines the common terms used throughout this paper.

\begin{table*}
    \fontsize{9pt}{9pt}\selectfont
        \centering
        \caption{Definitions of common terms and symbols used throughout this paper.}
        \label{tab:defs}
        \begin{tabular}{ll} 
                \hline
                Term &  \\
                \hline
            BCG (brightest cluster galaxy) & The BCG is the brightest galaxy in a cluster that lies close to the cluster centre. \\
                ICL (intracluster light) & Diffuse light that is not bound to any individual galaxy in the cluster but instead to the cluster potential\\
                Surviving component & The component of the ICL and BCG that was accreted from still surviving galaxies\\
                Completed mergers component & The component of the ICL and BCG that was accreted from galaxies that have been completely disrupted\\
                In situ component & The component of the ICL and BCG that formed in situ  \\
                Progenitor galaxy & The progenitor galaxies that formed the BCG+ICL. They may be totally disrupted or still surviving\\

                \hline
        \end{tabular}
\end{table*}

\subsection{Overview of simulations}
We used data from The Next Generation Illustris Project (IllustrisTNG; \citet{Marinacci2018,Naiman2018,Nelson2018,Nelson2019,Pillepich2018b,Pillepich2019,Springel2018}. The IllustrisTNG simulations are an improved version of the original Illustris simulations \citep{Vogelsberger2013,Genel_2014,Torrey_2015,Nelson_2015} with more realistic galaxy stellar masses and sizes \hl{\mbox{\citep{Genel_2017}}}, along with galaxy groups and clusters with gas fractions that are in better agreement with observations, while significantly improving  on the colour bimodality \citep[e.g.][]{Pop2022}. TNG uses the moving-mesh code \textsc{AREPO} \citep{Springel_2010} to follow the gas dynamics coupled with dark matter using a quasi-Lagrangian treatment within the $\Lambda$ cold dark matter paradigm. Dark matter halos in the simulation are identified using a friends-of-friends (FoF) algorithm (\citealt{Davis1985}), which selects groups of particles separated by a linking length equal to or smaller than 0.2 times the mean inter-particle separation. FoF groups are generally referred to as halos. Secondly, the \textsc{subfind} algorithm \citep{Springel2001,Dolag2009} is used to identify subhalos as gravitational-bound substructures within FoF halos. Throughout this paper, $M_{200}$ denotes the mass measured within the halo radius, $R_{200}$, that encloses an average over-density of 200 times the critical density of the universe, while {\mbox{$M^{*}_{200}$} refers to the stellar mass within the same radius.} Whether a cluster is relaxed or unrelaxed is defined by the 3D offset between the position of the BCG (defined by the most bound particle) and the centre of mass of the gas. A cluster is defined as disturbed if the offset between the BCG and the centre of mass is larger than $0.4\times R_{200}$ and relaxed if $D_{\text{BCG-CM}}$ {\mbox{($<0.1$ $R_{200}$;}} \citealt{Aldas_2023}).

The simulation includes boxes of 50, 100, and 300 Mpc per side, corresponding to TNG50, TNG100, and TNG300. Each of them consists of 100 snapshots from z = 20 to z = 0. Twenty snapshots are full, while the remaining eighty are limited. Both encompass the entire volume, but mini snapshots have only a subset of data available. Each box runs three simulations at different resolution levels. TNG includes new physical processes, such as magnetohydrodynamics, as well as prescriptions for black hole formation, growth, and multimode feedback, stellar evolution, and chemical enrichment \citep{Pillepich2018b}. Small galaxies are comparable with observations,  \hl{at a high level of resolution}. TNG uses cosmology parameters derived from The \citet{Planck2016} in which $\Omega$\textsubscript{m} = 0.31, $\Omega$\textsubscript{$\Lambda$}=0.69 $\Omega$\textsubscript{b} = 0.0486 and h = 0.677.

In this work, we used TNG100 because it includes a population of high-mass galaxy clusters balanced with a higher resolution than TNG300. \hl{IllustrisTNG-100 hosts 127 clusters and groups above a mass of \mbox{10\textsuperscript{13}~\(\textup{M}_\odot\)}, and 10 clusters with mass \mbox{>10\textsuperscript{14}~\(\textup{M}_\odot\)}, ranging from \mbox{10\textsuperscript{13}-2.6~$\times$~10\textsuperscript{14}~\(\textup{M}_\odot\)}}.
TNG100 is the middle-resolution baryonic run of the IllustrisTNG suite and has a volume of approximately 110\textsuperscript{3} Mpc\textsuperscript{3} and 2 × 1820\textsuperscript{3} resolution elements, with an average mass per baryonic
particle of 1.4~$\times$~10\textsuperscript{6}~\(\textup{M}_\odot\), and dark matter mass with  7.5~$\times$~10\textsuperscript{6}~\(\textup{M}_\odot\) per particle.

\subsection{Selection of ICL}
\hl{To separate satellites from the main halo, the IllustrisTNG simulations first identify distinct dark matter halos using FoF grouping and then use the \textsc{subfind} algorithm to find self-bound structures within the halo. The most bound sub-halo is identified as the central sub-halo. }All the star particles in a halo that are not dynamically associated with a particular galaxy are defined as being attached to the central sub-halo. Thus, while observations have to contend with the separation of satellite galaxies from the ICL, the primary concern in TNG100 is dividing the central BCG from the ICL. 

To begin with, before the division of the ICL and the BCG is performed, we rotated galaxies using the inertia tensor to ensure that we select galaxies with the same orientation. Plotted gradients of the 1D surface brightness, colour, and metallicity were performed using the semi-major axis. As previously discussed, the separation of the ICL from the BCG is one of the most difficult questions for both theoretical and observational studies, as there is often no strict way of dividing the two. As such, for some pieces of analysis, we  considered the BCG+ICL as a single system. For others, we performed a cut that strictly divides the ICL and the BCG. 

\begin{figure}

        \includegraphics[width=0.95\columnwidth]{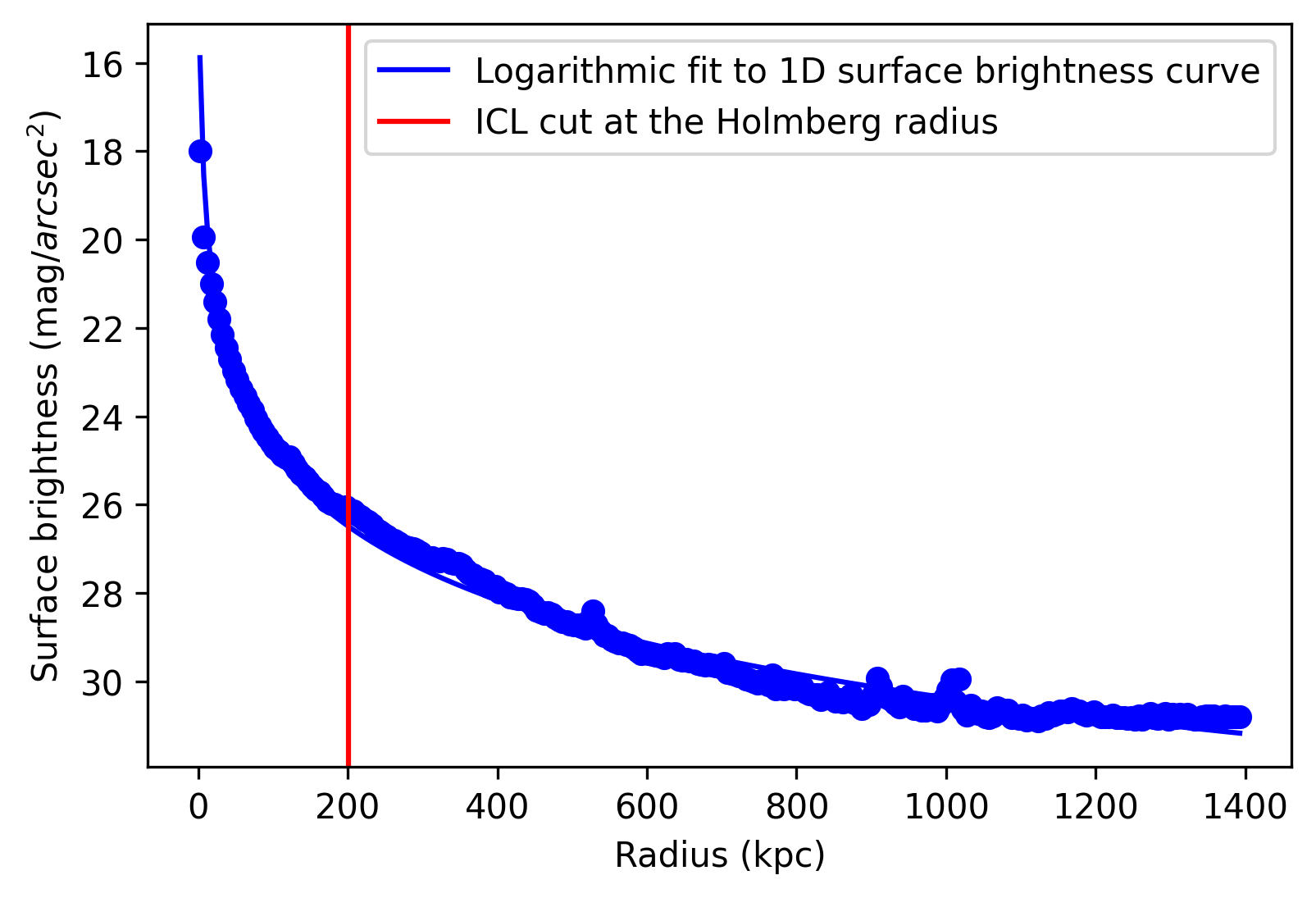}
    \caption{1D r-band surface brightness profile of the BCG+ICL of a cluster in TNG100 along the semi-major axis, with the red line indicating the division between BCG and ICL.}
    \label{fig:ID01D}
\end{figure}

To divide the ICL and the BCG, we created a 1D r-band surface brightness plot and use a surface brightness cut at the Holmberg radius \citep{holmberg1958} of 26.5 mag/arcsec\textsuperscript{-2}, where star particles within this radius are defined as being attached to the BCG and particles outside this radius are part of the ICL. Figure~\ref{fig:ID01D} shows an example of this for the dominant subhalo in one of our massive halos, with the red line showing the divide between the ICL and BCG. While we strictly divide the BCG and the ICL, some studies have additionally characterised a transition region where the BCG and the ICL overlap. We note that in this study, we did not attempt to characterise this region. A secondary issue is that we divide the ICL and BCG using a circular radius cut, which can be affected by the shape of the BCG. \hl{We also note that our chosen surface brightness cut may result in the loss of what other definitions would consider as ICL for particularly massive galaxies. In \mbox{Fig.~\ref{fig:ID01D}}, for example, the ICL cut is at 200 kpc, while other studies have separated the ICL and the BCG at radii such as 30 kpc \mbox{\citep[e.g.][]{Montenegro_Taborda_2023}} or 100 kpc \mbox{\citep[e.g.][]{Brough2023}}}. In the discussion, we consider how this choice would affect our results.  The outer limit for our ICL selection is the cluster virial radius of $R_{200}$, chosen because some particles defined as being attached to the cluster can extend well beyond the expected cluster radius.

\begin{figure}
        \includegraphics[width=0.95\columnwidth]{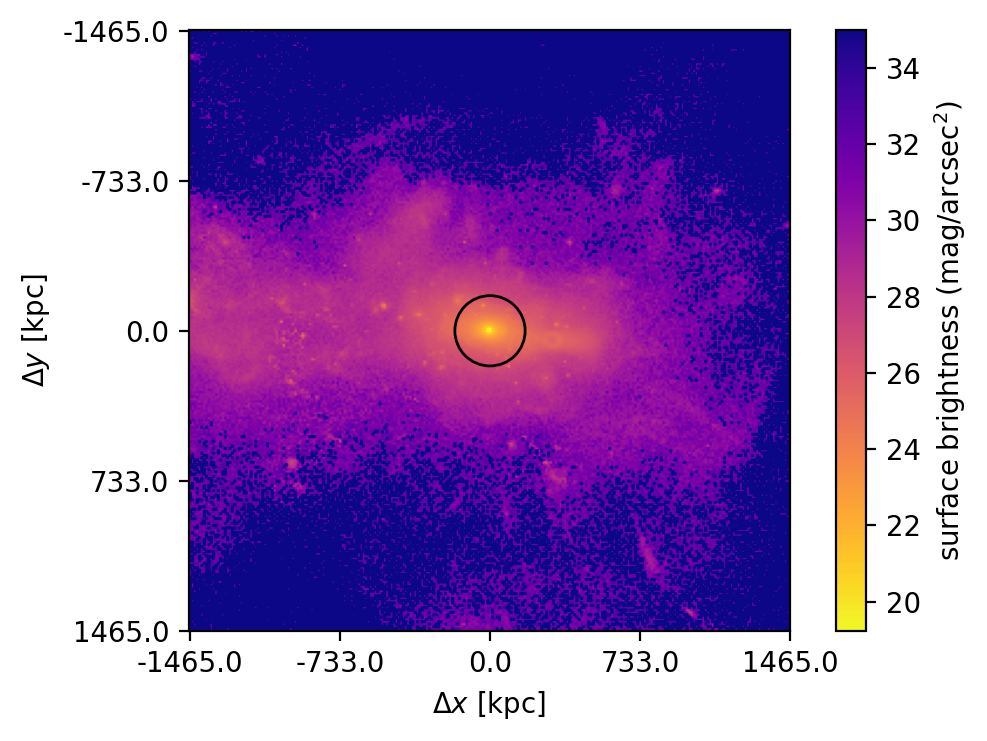}
    \caption{2D Surface brightness map of the BCG+ICL of a cluster in TNG100, centred on the BCG, with an upper limit of $R_{200}$. The black circle shows the divide between the ICL and the BCG. Light from the centre extends for hundreds of kiloparsecs and substructure is visible in the ICL. }
    \label{fig:ID02D}
\end{figure}

Figure~\ref{fig:ID02D} plots the 2D surface brightness of the BCG+ICL from the same cluster as in Fig.~\ref{fig:ID01D}, showing that light from the centre extends for hundreds of kiloparsecs. The black circle denotes the divide between the particles that make up the central BCG and the ICL. There are also structures visible in the ICL that reveal how its formation and evolution are influenced by the growth of the subhalos and halos within the cluster. 

\subsection{Origins of star particles}

To determine the origin of the particles that make up the BCG and ICL, we made use of the stellar assembly catalogues described in \citet{Rodriguez_Gomez2016}, which tags every particle as in situ or ex situ and further divides ex situ particles between those that were accreted by mergers or stripped from a galaxy that has not yet merged. After selecting the ex situ star particles, we  determined the masses of the original galaxies these particles were a part of. This was done through the following steps:
\begin{enumerate}
    \item Track back past snapshots for all particles bound to the central subhalo at z = 0 and determine the snapshot in which each star particle is no longer attached to the central sub-halo.
    \item Determine which sub-halo each particle is within for the snapshots selected in step (1), where the particle is no longer attached to the BCG.
    \item Track the sub-halos that the particles are found in via step (2) back through earlier snapshots and determine the maximum mass of all origin subhalos.
\end{enumerate}

Once this process was complete, we obtained  a final sample of sub-halos that contributed to the growth of the ex situ component of the ICL and BCG, along with their masses. 
A small number (between 1 to 2 per cent) of the sub-halos particles were found to be in at (2) were discovered to be in sub-halos that are anomalous. This can occur because the \textsc{subfind} algorithm cannot always differentiate between genuine subhalos and fragments or clumps in already-formed galaxies. Because of this, we also selected a second sample of origin sub-halos with the first requirement that particles are selected in the latest snapshot where they are located outside $R_{200}$, rather than the latest snapshot in which they are no longer attached to the central galaxy. If particles were found to be in sub-halos defined by a simulation as anomalous as that performed at step (2), they were replaced with sub-halos that met the $R_{200}$ requirement. Particles in anomalous galaxies for both catalogues were excluded from the analysis of origin galaxies. \hl{The mass of progenitor galaxies is defined as the mass within twice the stellar half-mass radius (HMR).} We note that we considered the ICL in both the group and cluster environment, but for simplicity, we only refer to the diffuse light component as the ICL (and not IGL).

\section{ICL mass fraction compared with cluster properties}
\label{section:ICLmassfrac}
\begin{figure}

        \includegraphics[width=0.95\columnwidth]{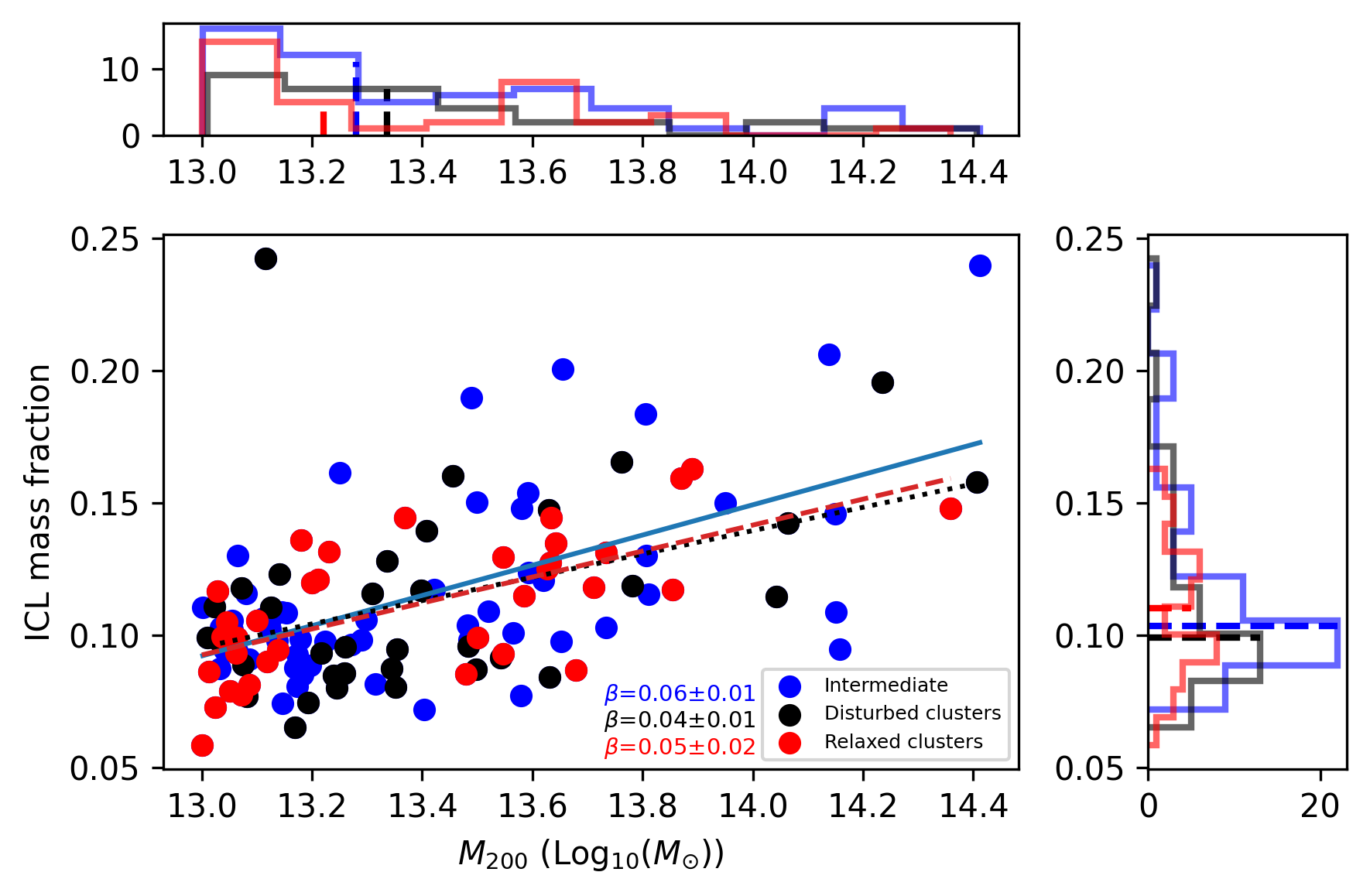}
    \caption{Mass fraction of ICL in galaxy clusters, separated by cluster dynamical state, with histograms of cluster masses and state. The dashed lines on the histograms indicate the median. The slopes of the linear fit to the data, $\beta$, are indicated in the bottom centre of the panel.}
    \label{fig:ICLmassfrac}
\end{figure}

In this section, we discuss the mass fraction that the ICL makes up of the host cluster and how it varies with properties such as cluster mass, concentration, and relaxation. Figure~\ref{fig:ICLmassfrac} plots the stellar mass fraction that the ICL makes up of clusters and groups against the total {stellar \mbox{$M^{*}_{200}$}} mass of the cluster. The relative fraction the ICL makes up of the stellar mass increases with mass. 
However, there is a lot of scatter in the relationship. Generally, the ICL makes up between 5 and 25 per cent of a cluster's stellar mass. The broad variation in ICL fraction at a given mass, consistent with that found in the literature \citep[e.g.][]{Contini_2021a}, is evidence of how the contribution of this component depends on both the physical mechanisms driving ICL fraction and halo properties linked to \hl{the} assembly history. 

\hl{Our results agree with results found in other studies, both theoretical and observational. In the simulations  \mbox{\citet{Ahvazi2024}} found ICL fractions of approx 9 - 14 per cent in TNG50, with the ICL selection within a radial range of \mbox{$0.15 r_{200} < r <  r_{200}$} from the host galaxy. \mbox{\citet{Pillepich2018b}} used fixed stellar apertures in IllustrisTNG and found ICL fractions of \mbox{$\geq$ 30} per cent. \mbox{\citet{Contreras_Santos_2024}} characterised ICL fractions in the Three Hundred Project with a 50 kpc aperture and found ICL mass fractions of 30–50 per cent within \mbox{$r_{500}$}. Observationally. \mbox{\citet{Mihos2017}} found an ICL fraction of between 7 and 15 per cent for the Virgo cluster, while \mbox{\citet{Furnell_2021}} found 20-40 per cent in observations of clusters, using a} {brighter} \hl{surface brightness B-band cut of \mbox{25 mag/arcsec\textsuperscript{-2}}. \mbox{\citet{Jim_nez_Teja_2018}} found ICL fractions of 2-23 per cent in 11 observed clusters, varying depending on cluster dynamical state,} {using three optical filters (F435W, F606W, and F814W)}. In general, ICL fractions in the literature can vary strongly, ranging from 5 to 50 per cent (\citealt{Contini_2021a}; see also \citealt{Tang2018} and \citealt{Kluge2021}) and are heavily dependent on how the ICL is selected \citep{Brough2023}. In particular, there is a great deal of variation, depending on how the border between the ICL and the BCG is drawn and whether studies characterise the transition region \hl{\mbox{\citep{Kluge2021, Contini_2022, Brough2023}}}. 
When we compare our results to those of other studies, we consider how their ICL definition may result in differing results.

In Fig.~\ref{fig:ICLmassfrac}, we also explore how the contribution from the ICL varies depending on the dynamical state of the clusters. Here, the cluster dynamical state is defined based on the 3D offset between the BCG and the gas centre of mass. Clusters are disturbed when the offset is larger than $0.4 \times \text{R}_{200}$, relaxed when the offset is {\mbox{$<0.1$ R$_{200}$}}, {and in an intermediate state when the offset is between \mbox{$0.4 \times \text{R}_{200}$} and \mbox{$0.1$ R$_{200}$}} \citep{Aldas_2023}. We find no clear correlation between cluster relaxation and the ICL mass fraction. Our lack of correlation could be a result of our definition of ICL mass fraction as being dependent solely on the stellar component of the BCG+ICL, with\hl{ no input from factors such as \mbox{$r_{200}$} or the velocity dispersion}. 

There are varying definitions of the dynamical state in the literature. Other examples include average sub-halo mass fraction \citep[e.g.][]{Jiang_2017,Cui_2018}, how well a halo obeys the virial ratio \citep[e.g.][]{Shaw_2006,Haggar_2024}, the mass difference between the BCG and the second most massive galaxy \citep[e.g.][]{Raouf_2019}, and other aspects. This is significant as different methods may measure different types of cluster dynamical states \citep{Haggar_2024}. Notably, our method is dependent on the gas and dark matter components of the cluster and independent of the stellar component that makes up the ICL. Our ICL \hl{definition} is also dependent on the BCG stellar structure and independent of our definition of dynamical state. Relaxation criteria that consider the stellar component of a cluster may be more likely to show a correlation either way.

\begin{figure}

        \includegraphics[width=0.95\columnwidth]{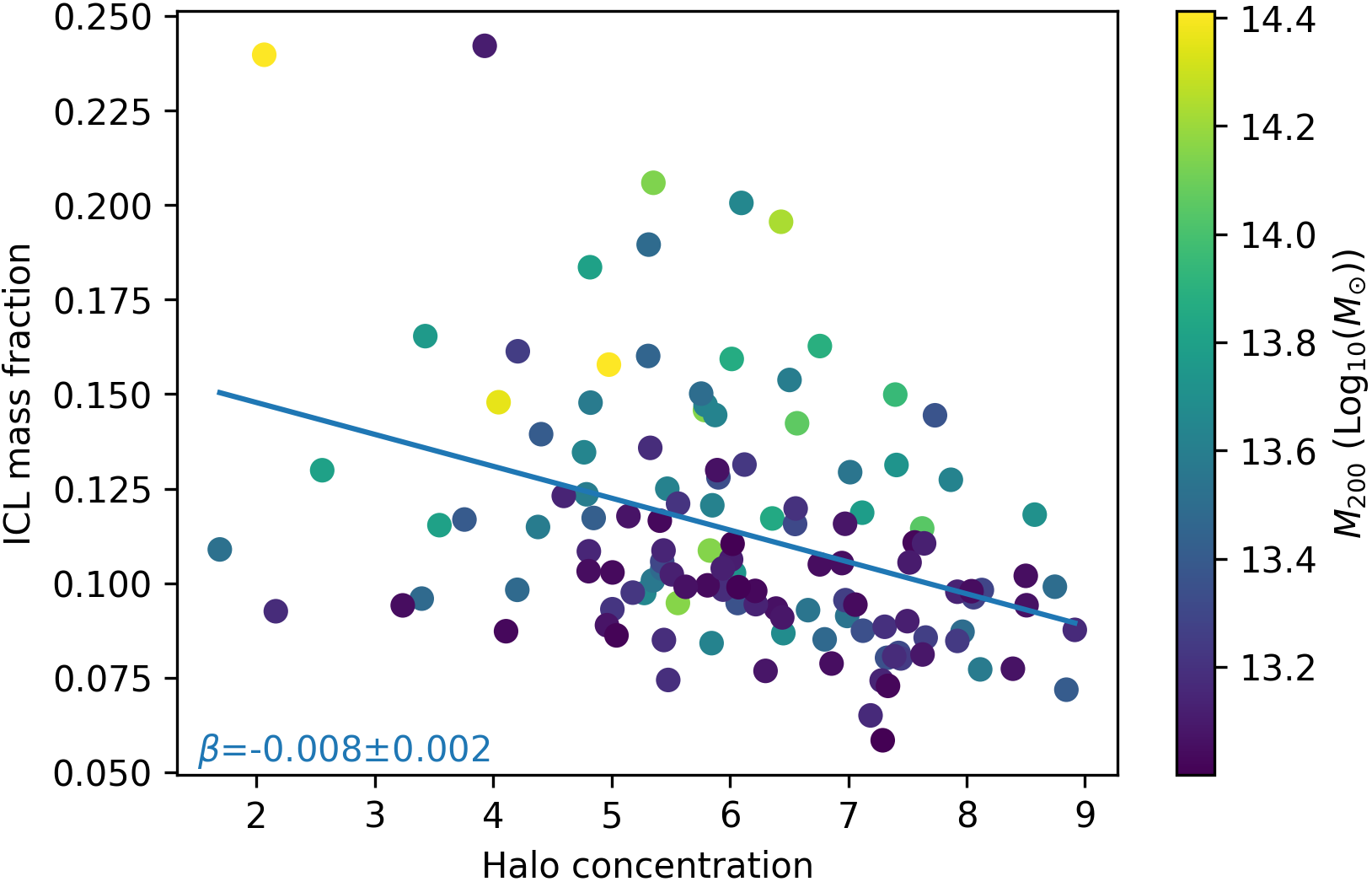}
    \caption{Mass fraction of ICL plotted against the cluster concentration, coloured by the cluster mass. The slope of the linear fit to the data, $\beta$, is indicated in the bottom-left of the panel.}
    \label{fig:conc}
\end{figure}

Figure~\ref{fig:conc} plots the ICL fraction against cluster concentration. Halo concentration in TNG100 is defined as $c=R\textsubscript{200c}/R\textsubscript{s}$, where $R\textsubscript{s}$ is computed by fitting an 
Navarro-Frenk-White profile to the dark matter density profile \citep{Anbajagane_2021}. \citet{contini2023a} found that more concentrated halos have a higher ICL fraction, suggesting that more dynamically evolved halos tend to have a higher proportion of ICL than less concentrated counterparts. More highly concentrated halos have stronger tidal forces and massive satellites experience faster dynamical friction, leading to more stellar stripping. In contrast, we find a weak decline in ICL fraction towards a higher concentration. 

However, at a given cluster concentration, higher-mass clusters appear to have higher ICL fractions, suggesting that controlling for concentration could allow for more accurate predictions of ICL mass fraction than cluster mass alone. Later in this work, we consider how concentration may impact the formation of the ICL by considering different ICL components and progenitor masses. Our different concentration results may be induced by different ICL selection methods. \citet{contini2023a} used a semi-analytic model to process the formation of ICL in galaxy clusters. In contrast, we selected ICL in fully formed clusters based on the surface brightness profile of the BCG, which is influenced by cluster concentration. Highly concentrated clusters may have more central stellar material and, therefore, the BCG may make up a larger fraction of the BCG+ICL system, meaning the ICL fraction will decline with concentration.

\begin{figure}

        \includegraphics[width=0.95\columnwidth]{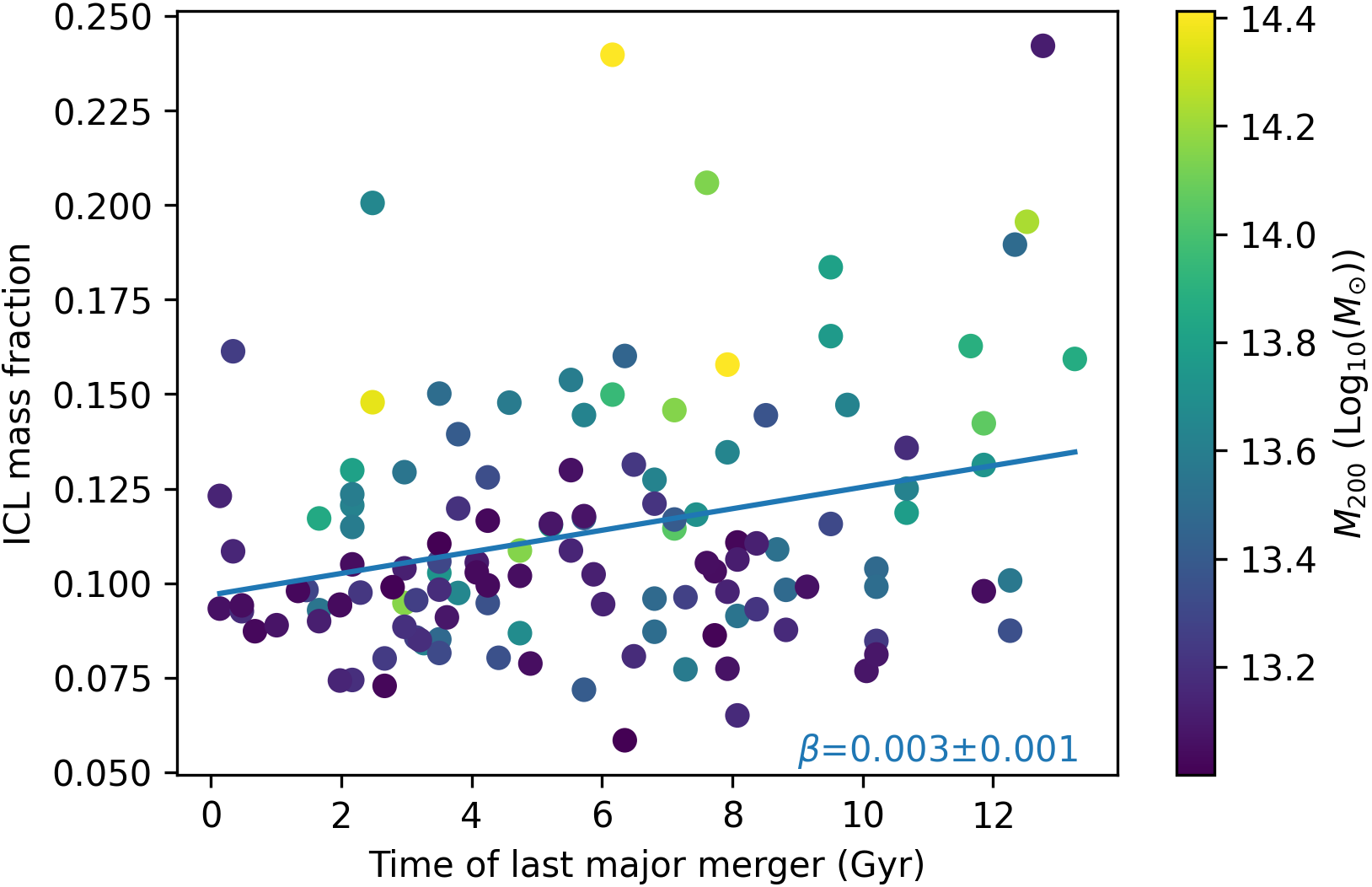}
    \caption{Mass fraction of ICL plotted against the lookback time since the last major merger, coloured by cluster mass.The slope of the linear fit to the data, $\beta$, is indicated in the bottom right of the panel.}
    \label{fig:massfracmajmerg}
\end{figure}

Figure~\ref{fig:massfracmajmerg} plots the stellar mass fraction of the ICL against lookback time since the BCG has undergone a major merger (defined as stellar mass ratio > 1/5). The ICL mass fraction increases with time since the last major merger, such that a more recent major merger indicates a lower ICL mass fraction. At a given merger time, more massive \hl{clusters} have higher ICL mass fractions.

The relationship between cluster relaxation, major mergers, and ICL fraction in the literature is uncertain. In their study, \citet{deOliveira2022} found that the total BCG+ICL mass fraction increases during major mergers, so that disturbed systems could show a higher ICL fraction. We note, however, that this applies to the full BCG+ICL system, rather than the ICL as a separate object, as in our study. Using simulations, \citet{Ahvazi2024} found that at a fixed halo mass, a longer time since the last major merger indicated a lower ICL fraction in TNG50. We note, however, that their definition of ICL fraction is based on the ratio \hl{between the ICL and} the BCG --- and not \hl{the ICL and} {stellar \mbox{$M^{*}_{200}$}} as in our study. Our differing results may indicate that major mergers have a greater impact on the growth of the BCG than the ICL. In contrast, another observational study by \citet{Ragusa_2023} observed a higher ICL fraction in groups and clusters with a greater proportion of
early-type galaxies, typically found in more evolved objects. These last results align with the theoretical predictions of \citet{contini2023a, contini2023b}, who found a higher average ICL fraction in halos that are more concentrated and formed earlier than others, \hl{a trend that} was later extended
to Milky Way-like halos \citep{contini2024}\hl{. More }relaxed systems may have had more time to have dynamical interactions, which results in an increased overall ICL fraction. \citet{Yoo_2024} used the cosmological hydrodynamical simulation Horizon Run 5 and found that an increase in the ICL fraction is not significantly impacted by \hl{cluster} merger events. At the time of the merger, they observed a substantial production of ICL, possibly due to preprocessed IGL being integrated into the cluster environment. However, the total ICL fraction displayed a slight downward trend, likely because \hl{a cluster} merger triggered star formation in \hl{satellite galaxies} \citep{Rudick_2011}. 
We conclude that we find some correlations between cluster properties and ICL mass fraction, but they show significant scatter. Moreover, we caution that these relationships may be heavily influenced by ICL selection criteria. Therefore, any comparison with observations must be carefully chosen.

\section{Components that make up the ICL and BCG}
\label{section:ICLBCGcomps}

The major ongoing question about the ICL and the BCG relates to how different components, such as stars stripped from surviving galaxies, stars accreted from completed mergers, and in situ stellar formation, contribute to the growth of both the ICL and the BCG. A follow-up question considers the masses and properties of the progenitor galaxies that stripped and merged to form the BCG and the ICL. In this section, we investigate the contribution of the three different components and how they are connected across the BCG+ICL system.

\begin{figure}

        \includegraphics[width=0.95\columnwidth]{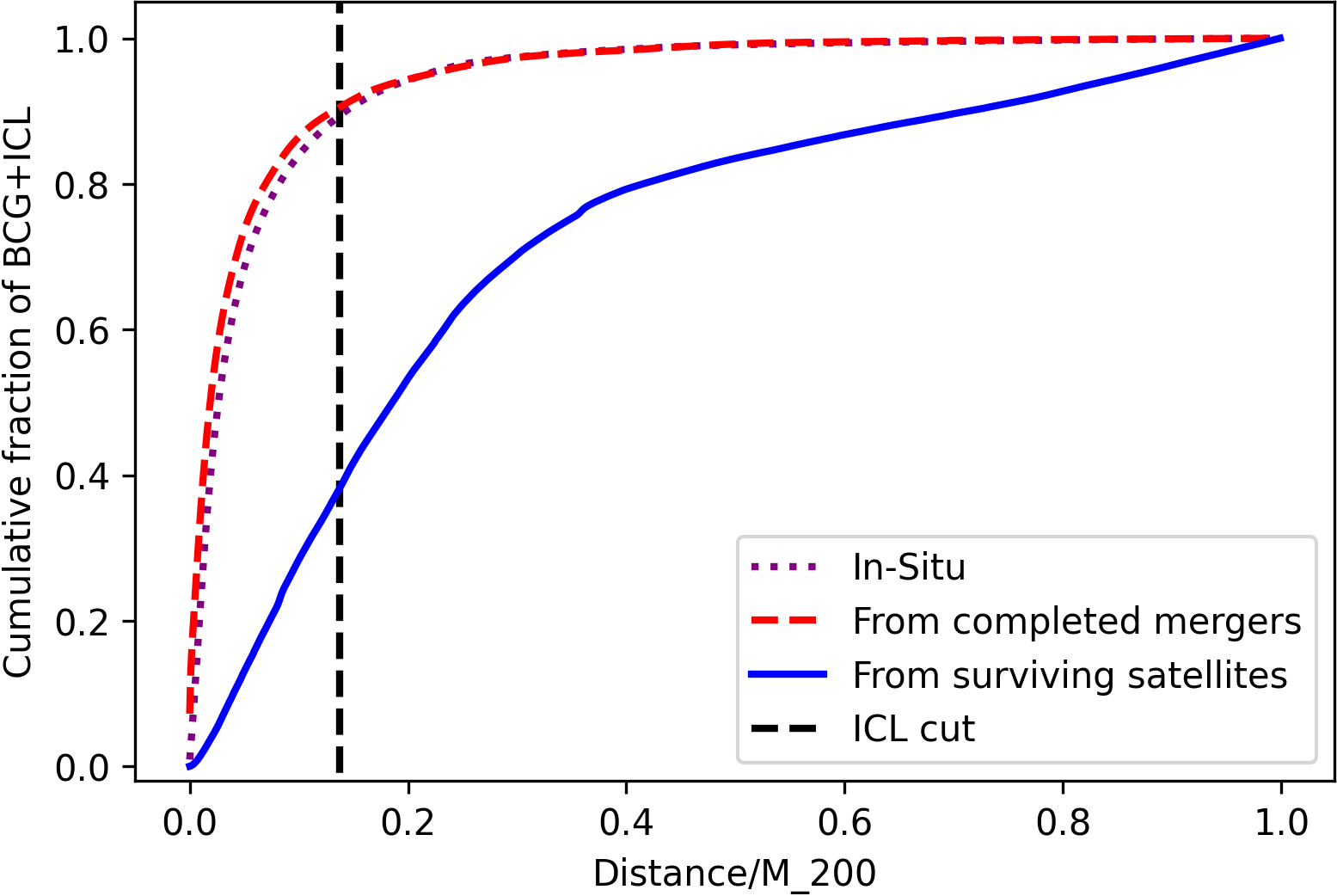}
    \caption{Radial distribution of the three different components of the BCG+ICL for an example massive cluster. The x-axis is normalised by $M_{200}$. The black dashed line indicates the division between the BCG and ICL.}
    \label{fig:ID0raddistcomps}
\end{figure}

Figure~\ref{fig:ID0raddistcomps} plots the radial distribution of the three different components of the BCG+ICL for the same massive cluster shown in Figs.~\ref{fig:ID01D} and~\ref{fig:ID02D}. We find that for this cluster, the component formed from stellar stripping is much more extended than the in situ and completed merger components, which are both centrally concentrated (with the in situ component being slightly more so), suggesting that for this cluster, the dominant contributor to the ICL is star particles from galaxies undergoing stellar stripping and ongoing mergers. The BCG is dominated by stars from in situ star formation and completed galaxy mergers.

\begin{figure}

        \includegraphics[width=0.95\columnwidth]{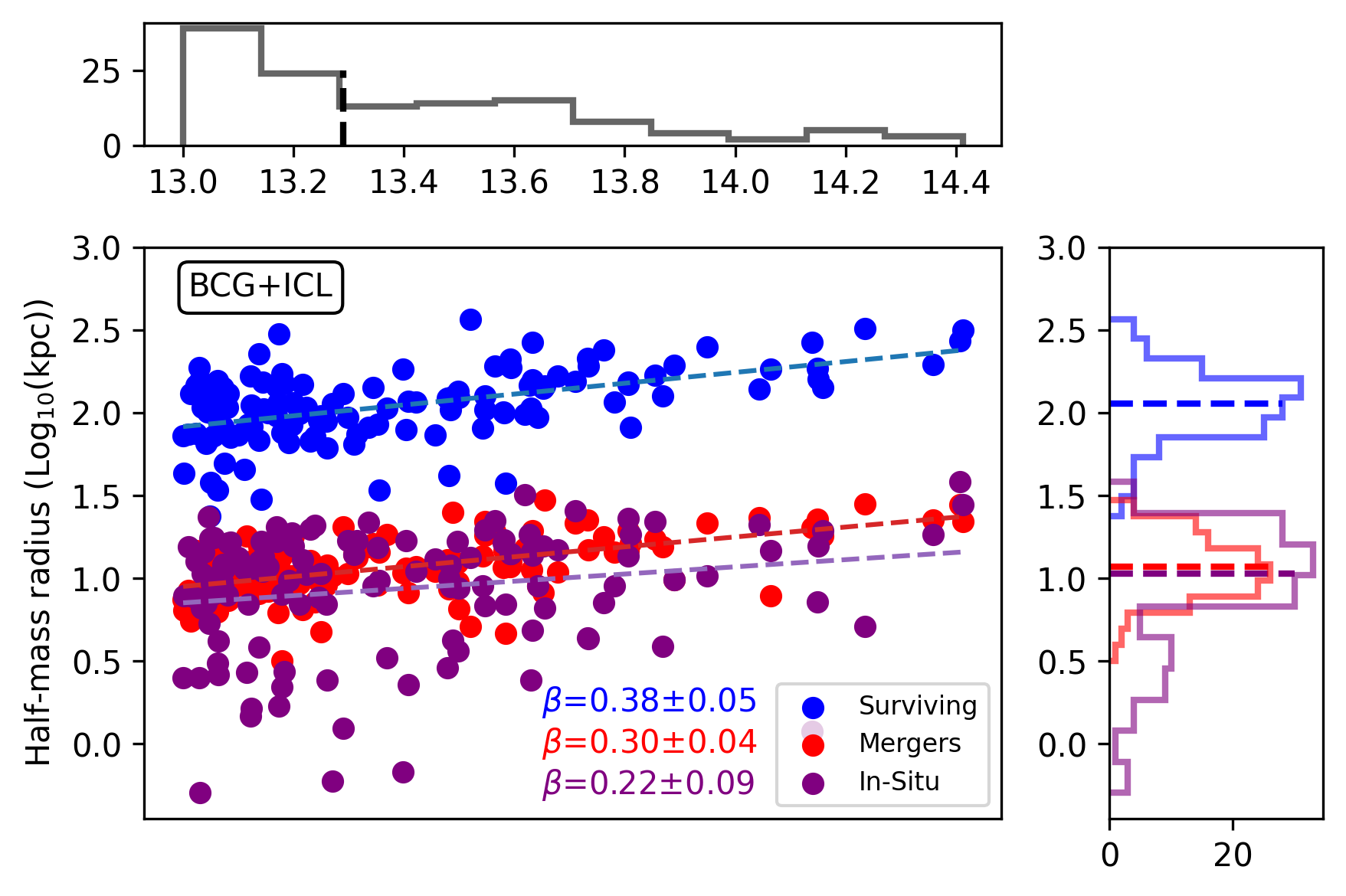}
    \includegraphics[width=0.95\columnwidth]{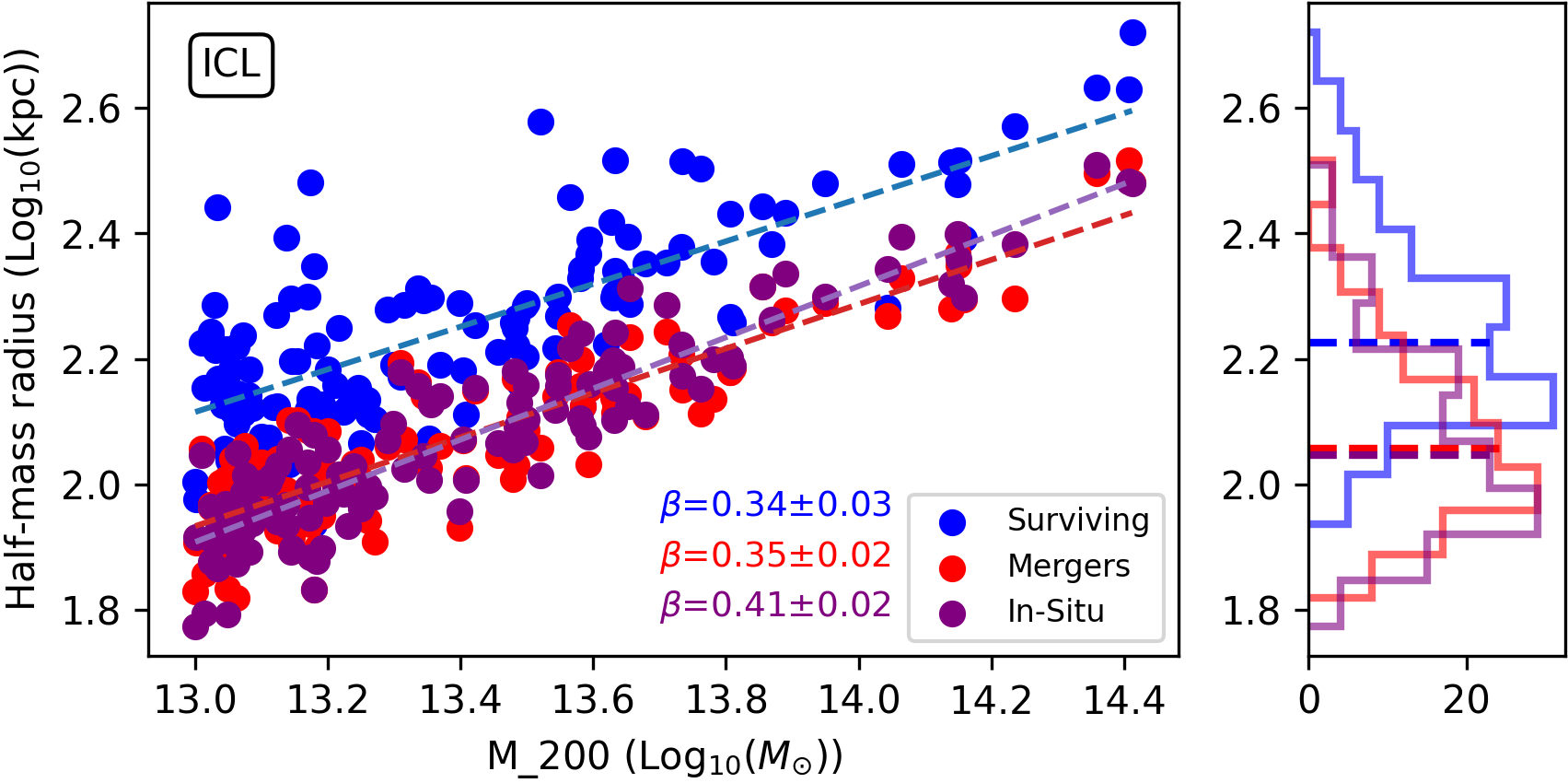}
    \caption{Top: HMR for the three different components that make up the BCG+ICL, the component accreted from surviving galaxies, accreted in completed mergers and formed in situ. Bottom: HMR for the three different components that make up the ICL. Dashed lines on the histograms indicate the median. The slopes of the linear fit to the data, $\beta$, are indicated in the lower middle panel.}
    \label{fig:compsHMR}
\end{figure}

The top panel of Fig.~\ref{fig:compsHMR} plots the HMR for the three different components that make up the BCG+ICL for all our clusters, against $M_{200}$. The most central component is star particles formed in situ, which overlap with the population formed from completed mergers. There is a greater scatter for the in situ population, likely due to individual variations in formation history. \hl{For example}, galaxies that have more recently experienced a major merger may be more likely to have in situ material in the outskirts of the galaxy due to disruption \hl{and ejection} \citep[e.g.][]{Khoperskov_2023}. \hl{Alternatively, it could be due to different formation histories.} The population formed from surviving galaxies is consistently more extended than the completed mergers and in situ components. 
The HMR of all three components increases with cluster mass. The bottom panel of Fig.~\ref{fig:compsHMR} plots the HMR for the three different components that make up the ICL for all our clusters, against $M_{200}$. The HMR of the components from surviving and merged galaxies have the same slope, indicating that they grow as part of the same process. Their slopes are also consistent with that for the full BCG+ICL system.
In contrast, the steeper slope of the in situ component, \hl{compared to the other two components that made up the ICL} indicates that the distribution of this component differs between the ICL and the full BCG+ICL system. 

Figures~\ref{fig:ID0raddistcomps} and~\ref{fig:compsHMR} suggest that a large reason for differences in the estimation of ICL mass fraction and composition across different studies is likely due to different methods of dividing the ICL and the BCG. Cuts such as the Holmberg radius, 100 kpc or three-component fitting will include less material in the ICL from completed mergers and in situ star formation than a more central cut of 30 kpc. At the same time, Fig.~\ref{fig:compsHMR} shows a great deal of individual variation in the distributions of the three different components between different clusters, suggesting that a universal radius cut is a flawed method of dividing the ICL and the BCG. \hl{Additionally}, selections based on dynamical information \hl{could} exclude merged or in situ material in clusters where these components are more extended. \hl{Alternatively, dynamical cut selections have been shown to put more material in the ICL than other measurements} \citep{Brough2023}. Therefore, when separating the ICL and the BCG, it is important to consider individual variations between clusters and how they may influence the results found.

We note that while Fig.~\ref{fig:compsHMR} shows the distribution of the three different components with mass and suggests that the more extended stripped component should dominate the ICL, it gives us no information about the masses of these three components and, thus, the relative amount each makes of the BCG+ICL system. 

\begin{figure}

        \includegraphics[width=0.95\columnwidth]{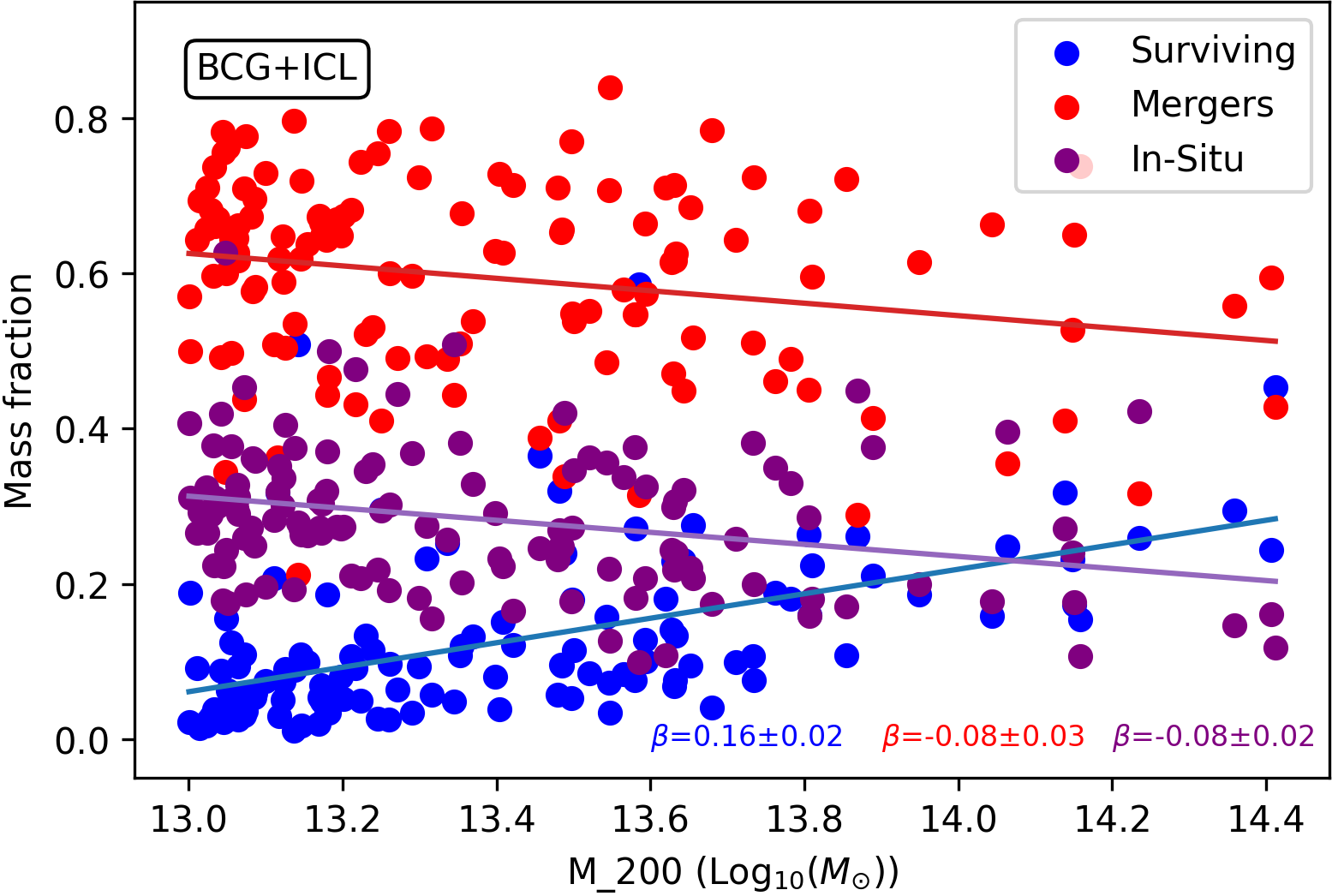}
    \caption{Mass fraction for the different components of the full BCG+ICL system. The component accreted from surviving galaxies, accreted in completed mergers, and formed in situ. The slopes of the linear fit to the data, $\beta$, are indicated in the bottom right of the panel. }
    \label{fig:compmassradio}
\end{figure}

Figure~\ref{fig:compmassradio} plots the masses of all three components for the full BCG+ICL system. It shows that the overall largest component of the system is from mergers, which have an overall mean contribution of $59 \pm 13$ per cent. However, there is a significant amount of scatter in the contribution of the merger component, with fractions ranging from 21 to 84 per cent. The relation declines slightly with mass, indicating that at higher masses, mergers play a slightly less dominant role in the formation of the full BCG+ICL system. The second most common component is the in situ component, which has an overall mean contribution of $28 \pm 9$ per cent and declines with mass similarly to the merged galaxies component. The smallest component is that of star particles stripped from surviving galaxies, which has an overall mean contribution of $12 \pm 10$ per cent. This component has the largest slope of the three relations, with more massive galaxy clusters having a much larger component of stars that have been stripped from surviving galaxies than smaller clusters. \citet{Pillepich2018b} investigated the stellar mass content of groups and clusters of galaxies in TNG300 and found that BCGs are made of ex situ material by more than 60 per cent even within their innermost 10 kpcs, in agreement with our results. \citet{Pulsoni_2021} studied the assembly of early-type galaxies in TNG100 and found that most profiles are dominated by in situ stars at small radii, but 8 per cent of objects are dominated by ex situ stars at all radii and these were the most massive objects.

\begin{figure}

        \includegraphics[width=0.95\columnwidth]{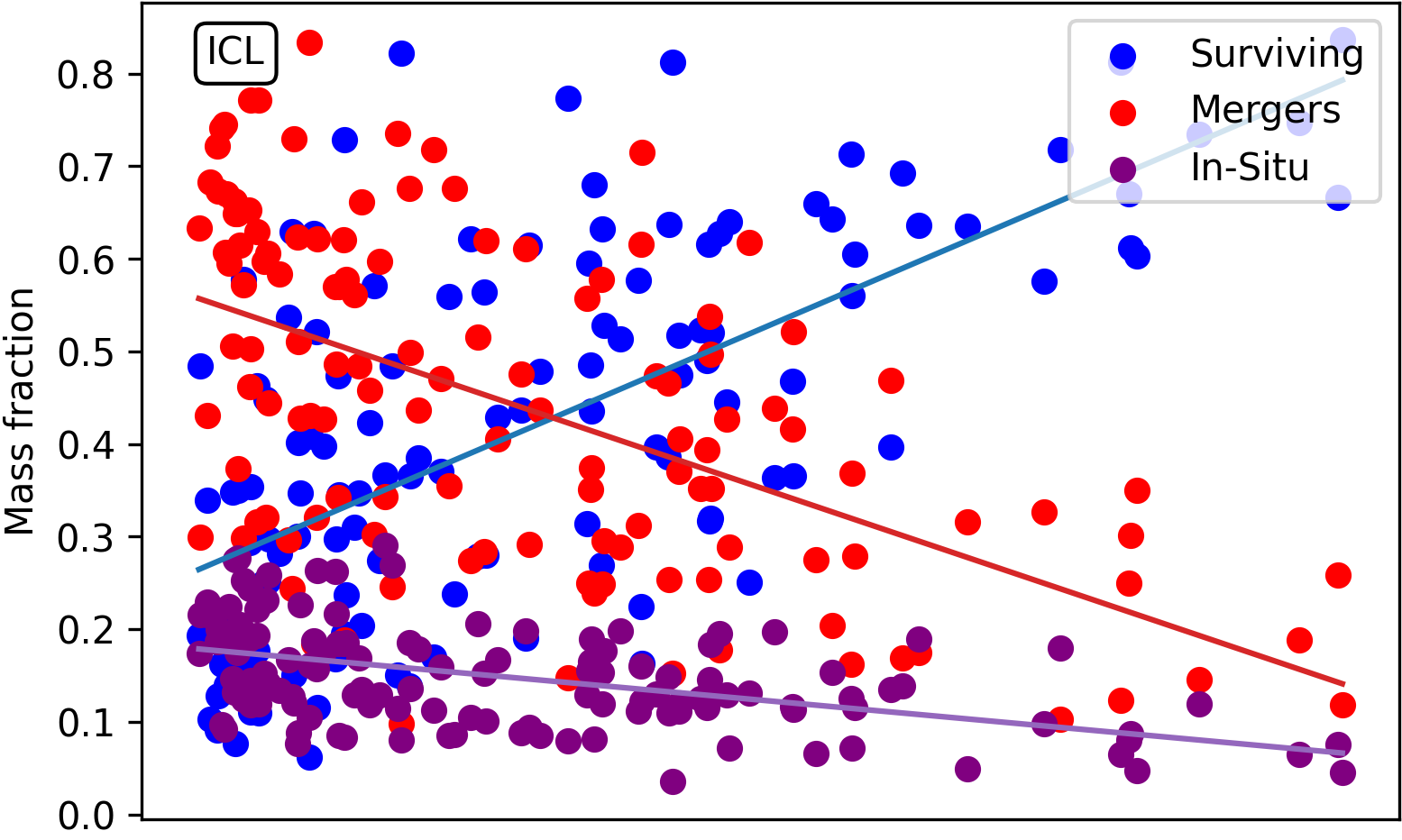}
    
    \includegraphics[width=0.95\columnwidth]{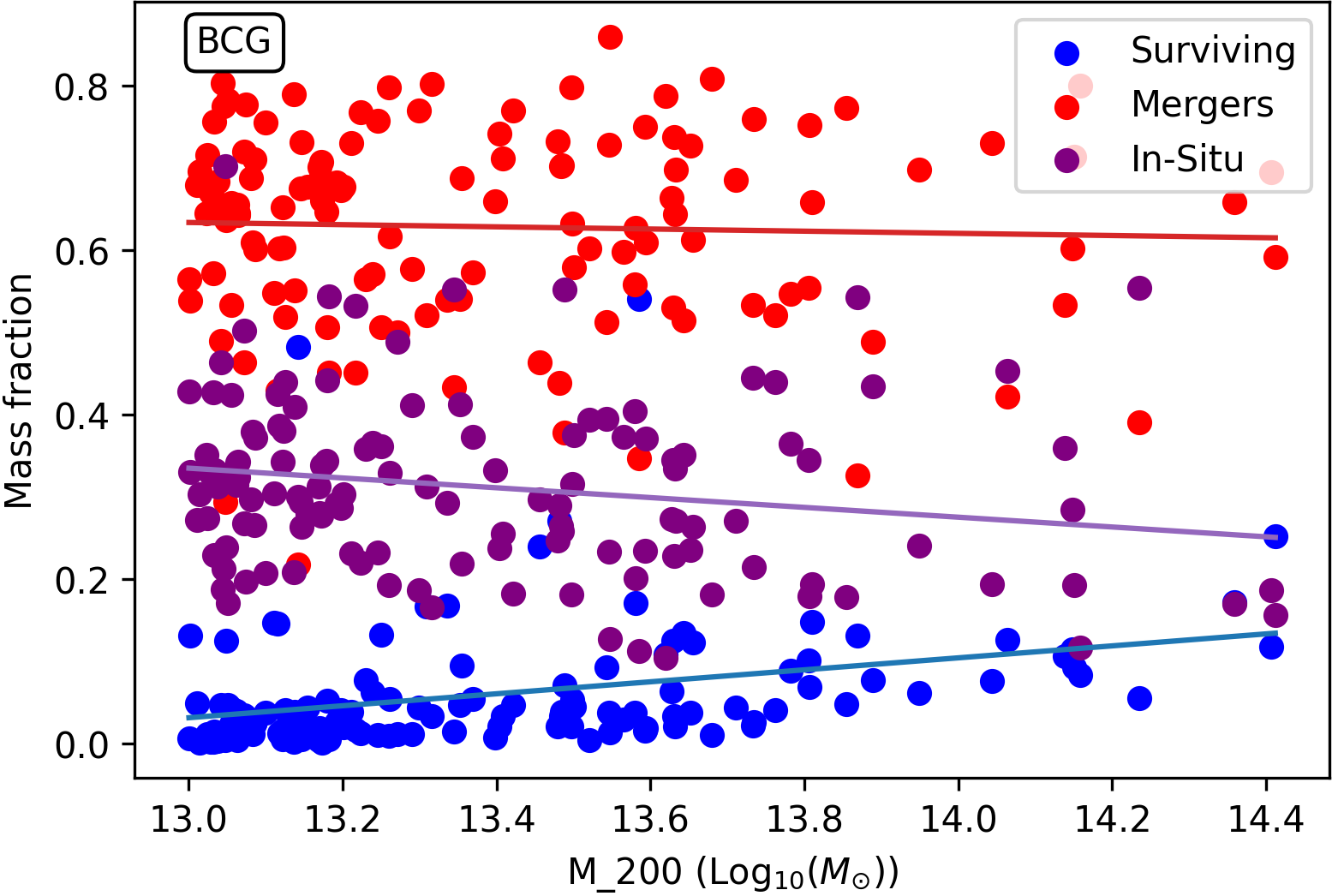}
    \caption{Top: Mass fraction for the different components that make up the ICL. The completed mergers component has $\beta = -0.29 \pm 0.03$. The in situ component has $\beta = -0.08 \pm 0.02$. The surviving galaxies component has $\beta = 0.37 \pm 0.02$. Bottom:  Mass fraction for the different components that make up the BCG. The completed mergers component has $\beta = -0.01 \pm 0.03$. The in situ component has $\beta = 0.06 \pm 0.03$. The surviving galaxies component has $\beta = 0.07 \pm 0.02$.}
    \label{fig:ICLmassratio}
\end{figure}

The top panel of Fig.~\ref{fig:ICLmassratio} shows the mass ratio specifically of the different components of the ICL. There are clear differences in fractions depending on the mass of the system. The component of the ICL stripped from surviving galaxies is the dominant component for high-mass clusters but declines to make up a much smaller fraction in the group environment, falling to as low as 10 per cent for the lowest-mass groups within our sample. Correspondingly, the component formed from merging galaxies increases with decreasing cluster mass. The two components become equal at a cluster mass of approximately $M_{200}$~$=$~3~$\times$~10\textsuperscript{13}~\(\textup{M}_\odot\). 

The decline in the component formed from completed mergers and the increase in that from galaxies that are stripped but not completely disrupted is likely a result of increased ongoing stripping in higher-mass environments. Satellite galaxies with higher masses cause and experience increased stripping in the outskirts of the clusters, forming parts of the ICL. Additionally, systems with more massive central galaxies will host more massive satellites, which will experience stronger dynamical friction than less massive galaxies. 

In situ stars in the ICL also decline with cluster mass, although at a much shallower slope. Notably, this plot shows that although the in situ component makes up the smallest component of the ICL, it is still fairly sizeable and increases with decreasing cluster mass, ranging from 10 to as high as 30 per cent in some lower-mass clusters. The relative increase towards lower-mass clusters and groups is likely partially a function of the decreased stripping in lower-mass environments. There is disagreement in the literature about the contribution of in situ star formation to the ICL. \citet{Puchwein2010} found that in situ star formation in simulations contributes about 30 per cent of the ICL.  \citet{Ahvazi2024b} studied in situ formation in ICL in three massive clusters in TNG50 and found 8 to 28 per cent formed directly within the ICL, hundreds of kiloparsecs from the centre of the host halo.  \citet{Montenegro_Taborda_2023} studied ICL in the \hl{TNG300} simulation and found that in situ material made up more than 10 per cent of the ICL. These authors suggested that this was due to either their chosen aperture of 30 kpc, major mergers pushing in situ stars towards outer regions, or the accretion of gas from satellite galaxies in the ICL. \citet{Melnick_2012} \hl{used population synthesis models to observe the ICL of an intermediate redshift cluster and found no more than 1 per cent of the ICL could be formed in situ}. \hl{To date, no observations have found ongoing in situ star formation in the ICL at z=0. } We find that even with larger aperture cuts in high-mass clusters, the in situ component still makes up around 10-20 per cent of the ICL. However, we also find that the in situ component of the ICL is the most centrally concentrated of the three different components that form the ICL and the BCG. \citet{Rodriguez_Gomez2016} studied the stellar mass assembly of galaxies in the Illustris simulation, and found that the accreted fraction of material in galaxies is strongly dependent on galaxy mass, with lower-mass galaxies having a larger in situ component. The decreased accretion for lower-mass galaxies is likely why lower-mass clusters have a larger relative in situ component in the ICL. \hl{We also note that we did not investigate the formation history of the in situ material in the ICL or whether it formed directly in the ICL itself or was distributed there by disruption, deferring this to a future study. In simulations of Milky Way-mass galaxies, in situ material has been found to make up to 30-40 per cent of the halo mass \mbox{\citep{Cooper2015Milky}.}}

The bottom panel of Fig.~\ref{fig:ICLmassratio} plots the mass fractions of the in situ, completed merger, and surviving galaxy components of the BCG (bottom). These plots show that there are clear differences in how the ICL and the BCG form. The in situ and completed merger components make up much greater fractions of the BCG and stay fairly consistent over the whole mass range, in contrast to the variation seen in the ICL. The slope of the in situ component of the BCG is consistent with the in situ relation of the ICL, suggesting that the in situ component of both the ICL and the BCG grows together and is less impacted by cluster mass and dynamics, although individual variations may still occur. The completed mergers component of the BCG follows a largely steady relation over the chosen cluster mass range, in contrast to the ICL component, which declines strongly with cluster mass.

The most notable difference is with the surviving galaxies component of the BCG. The component from surviving galaxies typically makes up an extremely small percentage of the BCG, with an overall mean of $6 \pm 8$ per cent. In comparison, the fraction of stars from surviving galaxies makes up an extremely large percentage of the ICL (over 50 per cent on average for clusters with mass greater than $M_{200}$~>~5~$\times$~10\textsuperscript{13}~\(\textup{M}_\odot\)). Combined with the plot of completed mergers, this suggests that the processes that drive galaxy stripping in the outskirts will impact how the ICL is formed, as well as whether the cluster is currently experiencing a major merger event. Note that this plot clearly shows that there are a small number of clusters where the BCG appears to have very large contributions from surviving galaxies. These BCGs are experiencing an ongoing merger with a massive galaxy. In all three cases, the strongest \hl{similarities} between the ICL and the BCG occurs in the low mass range, with distinct differences at higher masses, which is likely a consequence of ICL forming more efficiently in the cluster environment and being largely tied to stripping from surviving galaxies. \citet{Montenegro_Taborda_2023} investigated the ICL assembly of $M_{200}$~$\geqslant$~10\textsuperscript{14}~\(\textup{M}_\odot\) clusters in the TNG300 simulation and found a mean fraction of material stripped from surviving satellites to be 38.1 per cent. \citet{Brown2024} investigated ICL assembly in Horizon-AGN and found a stripped contribution of 44 per cent - this does not exclude ICL stars stripped from satellite galaxies that later merge with the BCG or otherwise do not survive until z = 0. \hl{These results are consistent with our finding that the stripped galaxies component makes up \mbox{41$\pm$20} per cent of the ICL, although our sample includes clusters of a lower-mass range.} \citet{contini2024} used a semi-analytic model and obtained a median z = 0 ICL contribution from {completed} mergers of only approximately 10 per cent for clusters
with halo masses between approx 10\textsuperscript{14}~\(\textup{M}_\odot\) and approximately 10\textsuperscript{14.5}~\(\textup{M}_\odot\). \hl{This is} {lower} \hl{than our range of \mbox{44$\pm$18}, but consistent with our clusters in the high mass range.}

\citet{Contini_2018} discussed that \hl{how different simulations define merger events can affect the} fractions of stellar stripping and merger material in the ICL. \hl{Numerical simulation and semi-analytic models can define the point of merger differently.} \citet{contini2014}, found that around 80 per cent of the ICL originates in stellar stripping, while \citet{Murante2007} found that around 75 per cent originates in mergers. This difference was found to be based on different definitions of when a merger occurred between the two studies. In semi-analytic models like \citet{contini2014}, a merger is defined as the point at which a satellite cannot be distinguished from the central body, while material accreted during the merger process is defined as originating from stellar stripping. In contrast numerical simulations such as \citet{Murante2007}, make estimates from merger trees. When \citet{contini2014} relaxed their definition of a merger, they found results consistent with \citet{Murante2007}. Therefore, caution must be used when quantifying the contributions of stellar stripping and mergers separately between different studies, as both originate from the same source, and differences often stem from variations in merger criteria.

\section{Correlation between ICL components and cluster properties}
\label{section:compcorr}
In this section, we explore the three different components that make up the ICL: material accreted from surviving galaxies, material accreted from completed mergers, and in situ star formation, along with their correlations with cluster state, concentration, and time of last major merger. \hl{We note that we did not factor in cluster mass, which may have an impact as the mass fraction these components make up of the ICL is dependent on mass.}
\subsection{Surviving component}

The mean stellar mass of the ICL surviving component is {\mbox{$41 \pm 20$} per cent}. To analyse the influence cluster history has on this component, we select clusters outside one standard deviation of the mean stellar mass fraction. This gives us 27 clusters with surviving mass $> 61$ {per cent} and 27 clusters with surviving mass $< 21$ {per cent}. Clusters with a high percentage of material from surviving galaxies in the ICL were more likely to be disturbed than those with a low percentage. 33 per cent (9/27) of the upper selection were disturbed, and only 15 per cent (4/27) of the lower selection. Conversely, the lower selection was slightly more likely to be relaxed. 33 per cent (9/27) of the lower selection were relaxed versus 22 per cent (6/27) of the upper selection. The number of clusters in an intermediate (\hl{in a transitional phase between disturbed and relaxed}) state were similar, 52 per cent (14/27) lower versus 45 per cent (12/27) upper. Clusters in a disturbed state are more likely to be undergoing active mergers, which leads to a larger amount of surviving material in the ICL.

Clusters with a high percentage of material from surviving galaxies in the ICL were found to have lower concentrations. The upper selection had a mean concentration of $5.2 \pm 1.2$, and the lower selection had a mean concentration of $7.0 \pm 1.1$. This is likely the reason why the ICL mass fraction \hl{in this component} is found to decrease with more concentrated clusters. More concentrated clusters are more evolved \citep{Sereno_2013}. Merging is more efficient in these clusters and, thus, more likely to contribute to the growth of the BCG than the ICL. While the concentration is tied to the growth of the ICL and BCG, the exact relationship may also depend on how the ICL is defined. 
The mean lookback time of the last major merger snapshot for ICL, with a high percentage of the surviving component, is $7.4 \pm 4.1$ Gyr, and for a low percentage, it is $4.2 \pm 2.7$ Gyr. This may indicate that a recent major merger results in a higher percentage of completed merger material being found in the ICL. However, the high uncertainties may be due to the fact that major mergers \hl{have not been found} to play a large role in ICL growth \citep{demaio2015}.

\subsection{Completed mergers component}

The mean stellar mass of the completed mergers component of the ICL is $44 \pm 18$ {per cent}; thus, we selected clusters outside of one standard deviation, giving 24 clusters with mass $> 62$ {per cent} and 24 clusters with $< 26$ {per cent}. 
Clusters with a high percentage of stellar mass coming from completed mergers in the ICL were less likely to be disturbed than those with a low percentage. 17 per cent (4/24) of the upper selection were disturbed, and 29 per cent (7/24) of the lower selection. Both selections were equally likely to be relaxed, with 29 per cent (7/24) relaxed clusters, while the upper selection was more likely to be in an intermediate state, 54 per cent (13/24), compared to 42 per cent (10/24). The relationship with concentration is the reverse of the surviving component. The upper selection had a concentration of $6.8 \pm 1.1$, and the lower selection, $5.3 \pm 1.2$. 
While the differences are within uncertainties, this suggests that more highly concentrated clusters are more evolved and, consequently, completed mergers make up a larger percentage of the ICL. The fraction with a higher percentage of completed merger material in the ICL had a more recent major merger, at $4.1 \pm 3$ Gyr ago. In contrast, the lower fraction had an earlier major merger $8.2 \pm 4.1$ Gyr ago. \hl{Clusters with a high fraction of completed mergers in the ICL are more evolved because merger material must be accreted at early times for progenitor galaxies to be entirely disrupted by z = 0. This would seem to contradict the time since the most recent major merger; however, note that this does not factor in the possibility that clusters with a low fraction of completed merger material are undergoing or are about to undergo a major merger, which is likely for a cluster which has not experienced a major merger for a long time \mbox{\citep[e.g.][]{Amorisco2017}}. }

\subsection{In situ component} 

The mean in situ stellar mass fraction of the ICL is $15 \pm 6$ per cent, so we select 19 clusters with mass $> 21$ {per cent} and 19 clusters with mass $< 9$ {per cent}. 
Clusters with a high percentage of in situ material in the ICL were more likely to be in an intermediate state, slightly less likely to be in a relaxed state, and less likely to be disturbed than those with a low percentage. In the upper selection, 5 per cent (1/19) were disturbed, 26 per cent (5/19) were relaxed, and 68 per cent (13/19) were intermediate. In the lower sample, 26 per cent (5/19) were disturbed, 21 per cent (4/19) were relaxed, and 53 per cent (10/19) were intermediate. Notably, clusters with a high fraction of in situ material in the ICL are considerably less likely to be disturbed than those with a high fraction of completed mergers and stripped material, which is likely a result of ongoing disruption resulting in the in situ material in the ICL being less significant. Clusters with a high percentage of in situ material in the ICL are more likely to be in an intermediate state, which is possibly due to in-situ material formed in the BCG being ejected during a past major merger. Clusters with a higher in situ fraction in the ICL were more highly concentrated. The upper in situ sample had a mean concentration of $6.6 \pm 1.3$, and the lower sample $4.8 \pm 1.4$, indicating that clusters with larger in situ fractions in the ICL are more evolved. The upper selection was less likely to have a more recent major merger than the lower fraction, although within uncertainties. The upper selection had a most recent major merger $6.5 \pm 3.2$ Gyr ago, and the lower at $5.5 \pm 3.5$ Gyr ago.

\section{ICL and BCG properties}
\label{section:ICLBCGprops}

In this section, we discuss the internal properties of the ICL and the BCG, their metallicities, colours, and ages.

\subsection{Metallicity and colour}

\begin{figure}
    \raggedleft
        \includegraphics[width=0.95\columnwidth]{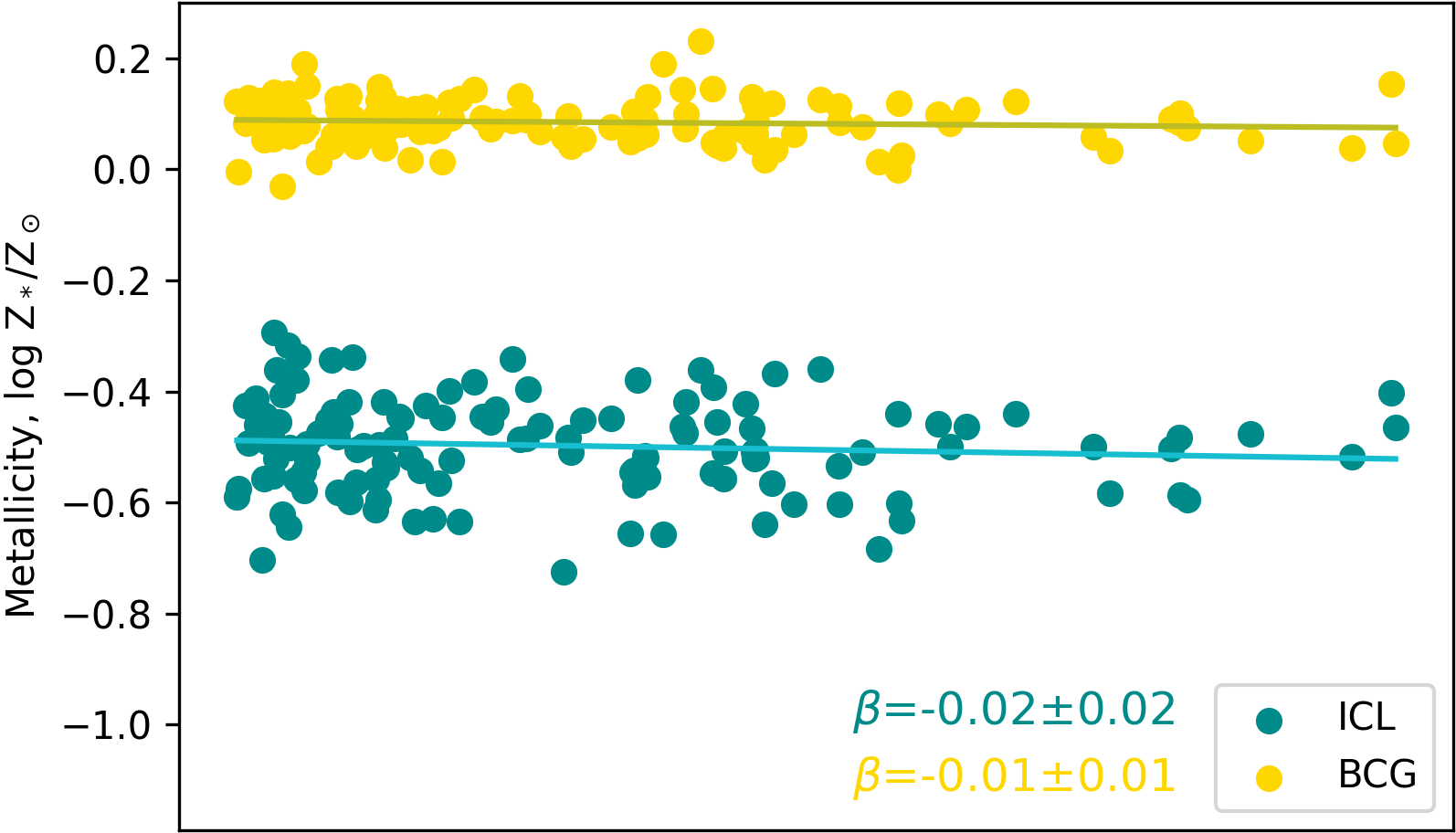}
    \includegraphics[width=0.95\columnwidth]{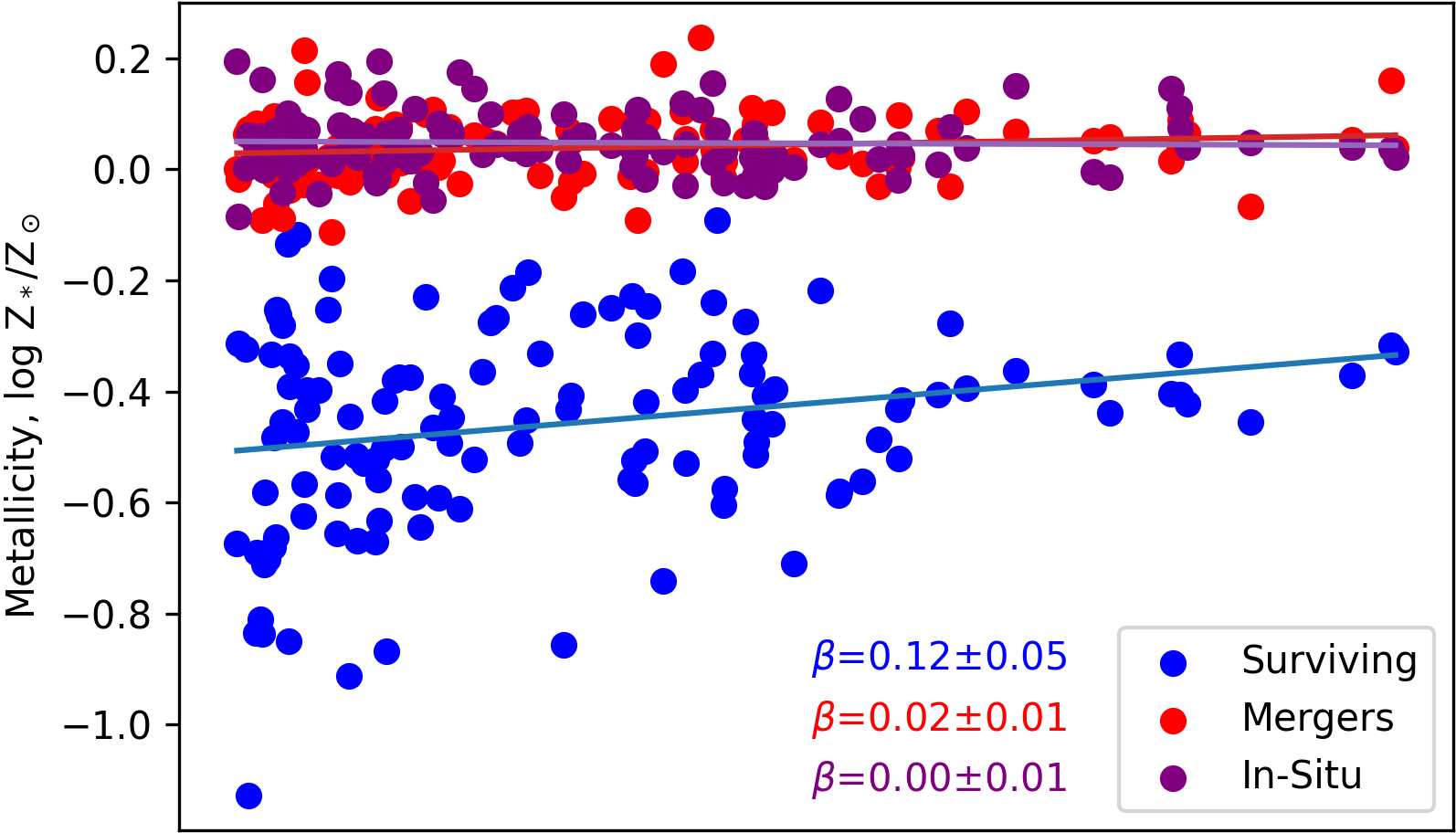}
    \includegraphics[width=0.95\columnwidth]{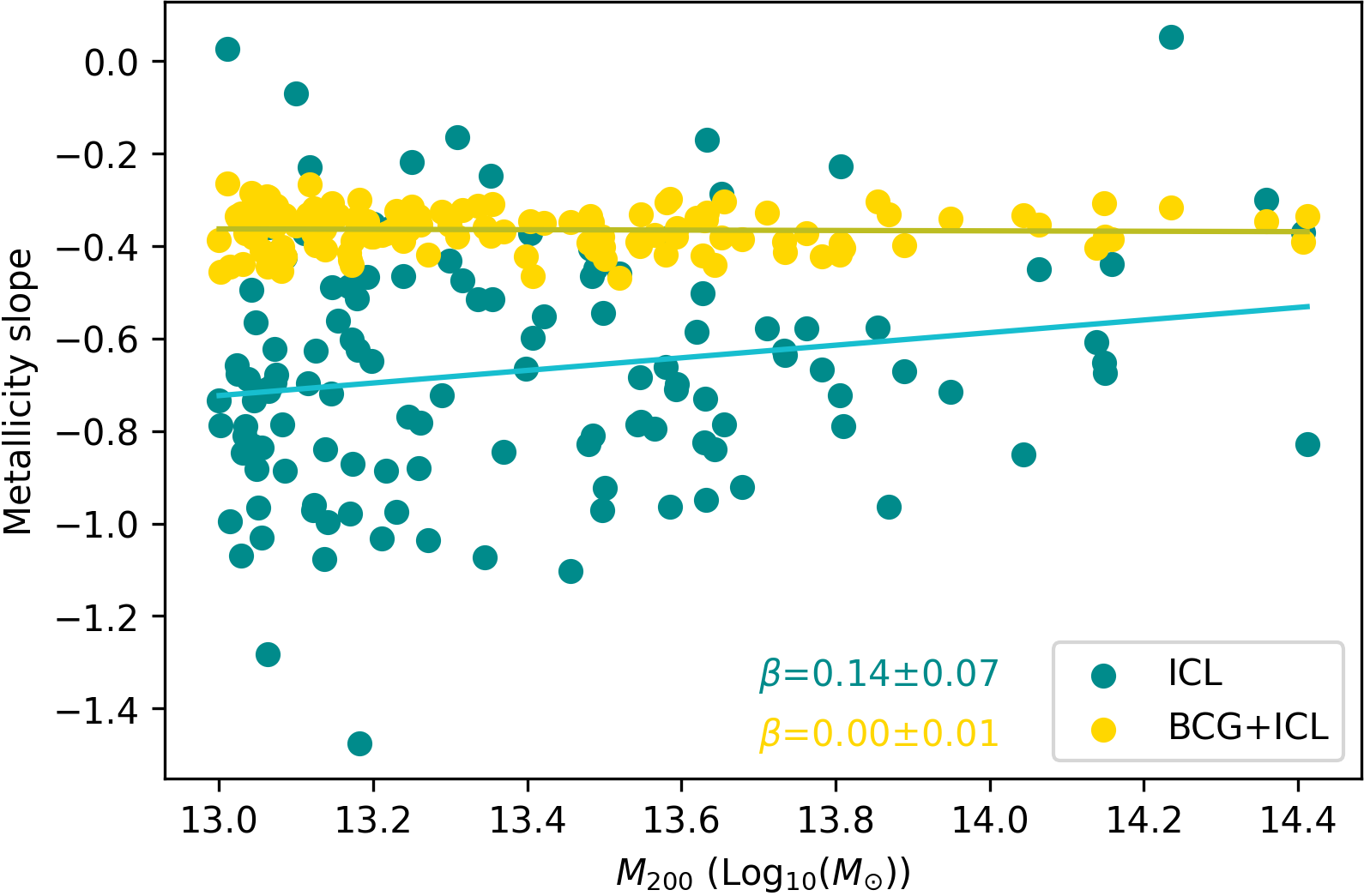}
    \caption{Top: Mean metallicity, log Z$_*$/\(\textup{Z}_\odot\) of the ICL and the BCG. Middle: Mean metallicity, log Z$_*$/\(\textup{Z}_\odot\) of the three different components of the full BCG+ICL system. Bottom: Cluster slopes of the metallicity of the BCG+ICL plotted over the full cluster radius, from the centre of the BCG out to $R_{200}$. The slopes of the linear fit to the data, $\beta$, are indicated in the bottom centre and right of the panel.}
    \label{fig:Z}
\end{figure}

The top panel of Fig.~\ref{fig:Z} plots the mean metallicity log(Z/\(\textup{Z}_\odot\)) of the ICL and the BCG of our simulated clusters. Over the mass range chosen, \hl{the mean} metallicity does not evolve with cluster mass. However, there is a significant offset in metallicity between the ICL and the BCG, with the ICL being considerably more metal-poor (by approximately 0.6 dex) and featuring slightly more scatter, which is likely a reflection of how ICL formation histories may vary between clusters. 
The middle panel of Fig.~\ref{fig:Z} plots the metallicity of the three different components that make up the BCG+ICL and finds that there is no significant difference in metallicity between the in situ and completed mergers components for the mass range chosen. 

The stellar mass stripped from surviving galaxies is significantly more metal-poor than either of the other two components. Thus, the ICL is more metal-poor because it is primarily composed of this component. There is considerably more scatter in the metallicity of this component. It is also the only relation with a notable slope, with more massive clusters being more metal-rich due to the accretion of more massive, metal-rich galaxies. 
The increase in scatter is likely due to individual variations between the types of galaxies being stripped in different clusters. In some clusters, the stripped component may be dominated by massive metal-rich galaxies in the centre of the cluster that will soon merge with the BCG. In contrast, in other clusters, the stripped component may be dominated by metal-poor galaxies that have experienced stripping on the outskirts of the cluster. Towards higher cluster masses, massive, metal-rich galaxies are more likely to experience stripping, resulting in the increased metallicity of this component with cluster mass.

The bottom panel of Fig.~\ref{fig:Z} plots the metallicity gradients of different clusters over the BCG+ICL, and the ICL region. This plot is made by selecting the star particles that make up the BCG+ICL over a 15-degree selection along the major axis, \hl{chosen because gradients over the major and minor axes can differ \mbox{\citep[e.g.][]{Monachesi2016}}. The star particles of the ICL+BCG are then divided into 30} bins from the centre of the BCG to the outskirts of the cluster, such that each bin contains the same number of particles. \hl{The mean metallicity within each bin is then plotted and a linear fit is applied to find the metallicity gradient of each cluster, on a logarithmic scale.} This is then repeated separately for the ICL region. The gradient of each of the 127 clusters is then selected and plotted. There is no distinct trend over $M_{200}$ for the BCG+ICL, but the ICL gradients increase with cluster mass. The scatter also increases considerably towards lower masses. This suggests that the trend in metallicity is heavily influenced by the individual formation history of the BCG+ICL. Notably, all clusters over the given mass range were found to have a negative metallicity gradient, ranging from -0.2 to -0.45 for the BCG+ICL, and 0.05 to -1.48 for the ICL, indicating that in all BCG+ICL systems, metallicity declines towards the outskirts, regardless of whether the ICL was formed due to stellar stripping or mergers of low- or high-mass galaxies. This is due to the metal-poor contribution from surviving satellites, which dominates at the outskirts. A small number of ICL systems have slopes close to flat or with slight positive slopes, possibly due to stripping of more metal-rich galaxies in the outskirts of these clusters.

Relaxation processes during mergers tend to mix up the stellar populations \citep{Murante2007}. When a satellite merges with the BCG, some of its stars become part of the ICL. As a result, the likely outcome of the ICL being formed in mergers is a constant colour-metallicity gradient regardless of distance from the centre. Some observational studies support this \citep{Ragusa_2021,Joo_2023}, while others support a colour gradient \citep{demaio2015,iodice2017,Morishita2017,Montes2018,Raj2020}. Observations generally find a negative gradient in colours or metallicity \citep[e.g.][]{demaio2015, montes2022}.  \hl{Some observations have revealed that gradients vary between clusters and that they can include flatter} distributions, {however, the gradients come with significant uncertainties} \citep{Krick_2007}.

\hl{Simulations have shown that} the most likely cause of a gradient is stellar stripping \citep{contini2024}. This mechanism is gentler and it is also likely that only the outermost stars of a satellite end up getting stripped. In contrast, the innermost stars will eventually become part of the BCG once the satellite completes the merger. Whether the gradient is positive or negative depends on the type of satellites undergoing stripping. Intermediate or massive satellites will have younger, bluer stars being stripped from their outskirts. \citet{Contini2019} suggested that the presence of gradients implies a primary, gradual formation mechanism of stellar stripping, while the absence of gradients suggests mergers are the primary formation mechanism. We find, however, that the presence of a negative gradient in metallicity and colour is universal regardless of whether ongoing or completed mergers are the dominant contributor to the ICL of individual clusters, with a high degree of active stellar stripping likely impacting the degree of the gradient.  
Stellar stripping can also induce flat distributions if the satellites being stripped are similar in colour and metallicity to the BCG.
\begin{figure}

    \raggedleft
        \includegraphics[width=0.98\columnwidth]{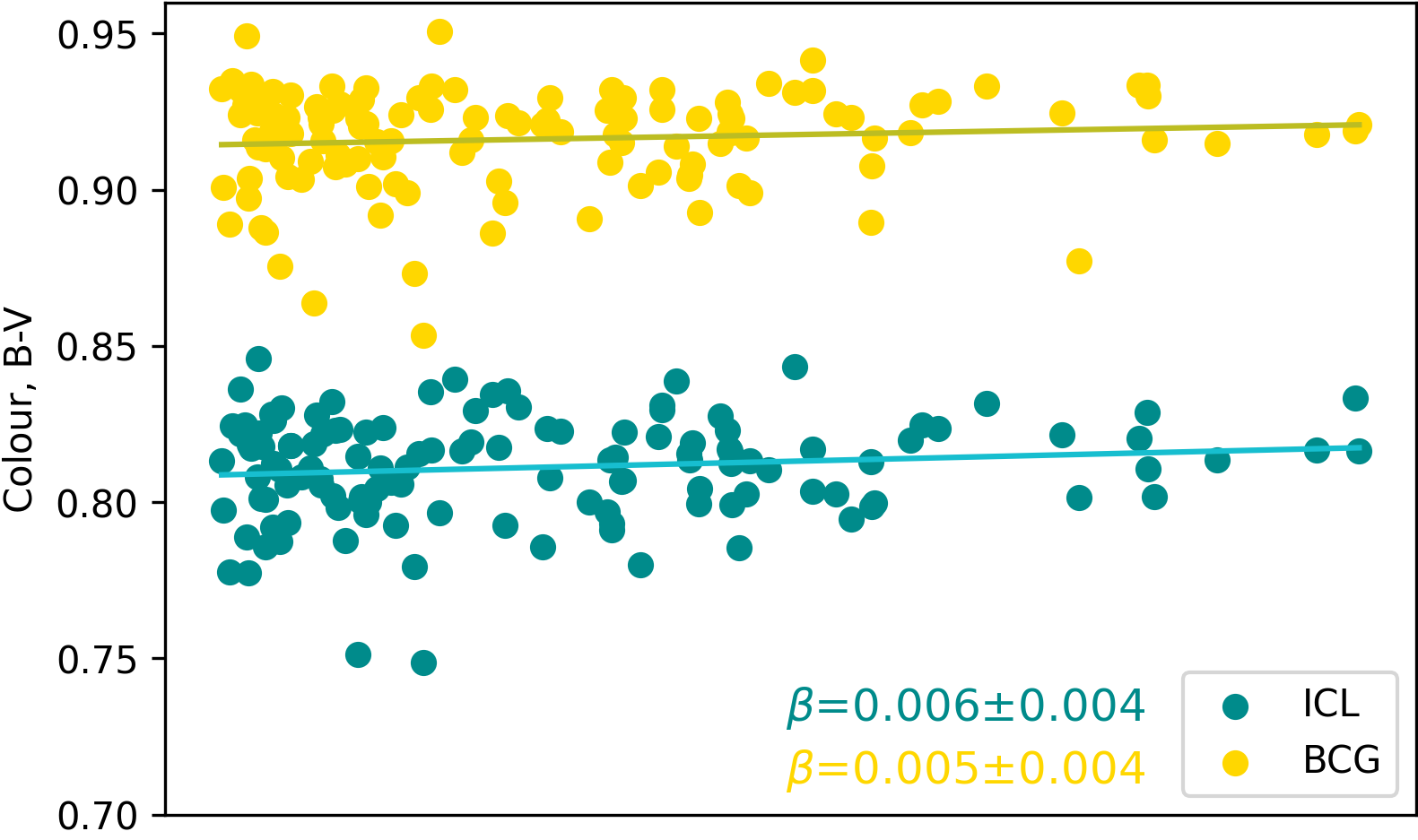}
    \includegraphics[width=0.98\columnwidth]{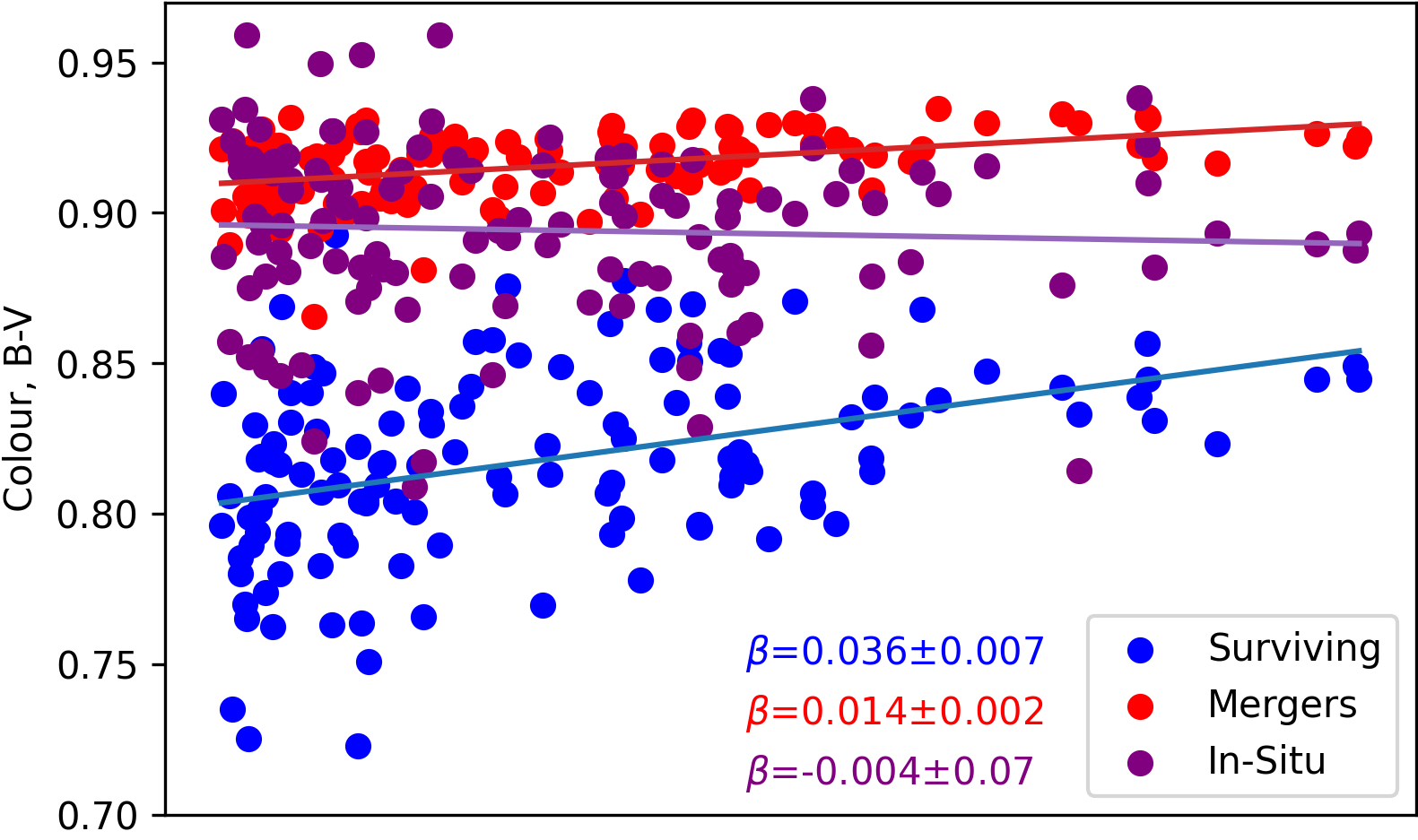}
    \includegraphics[width=\columnwidth]{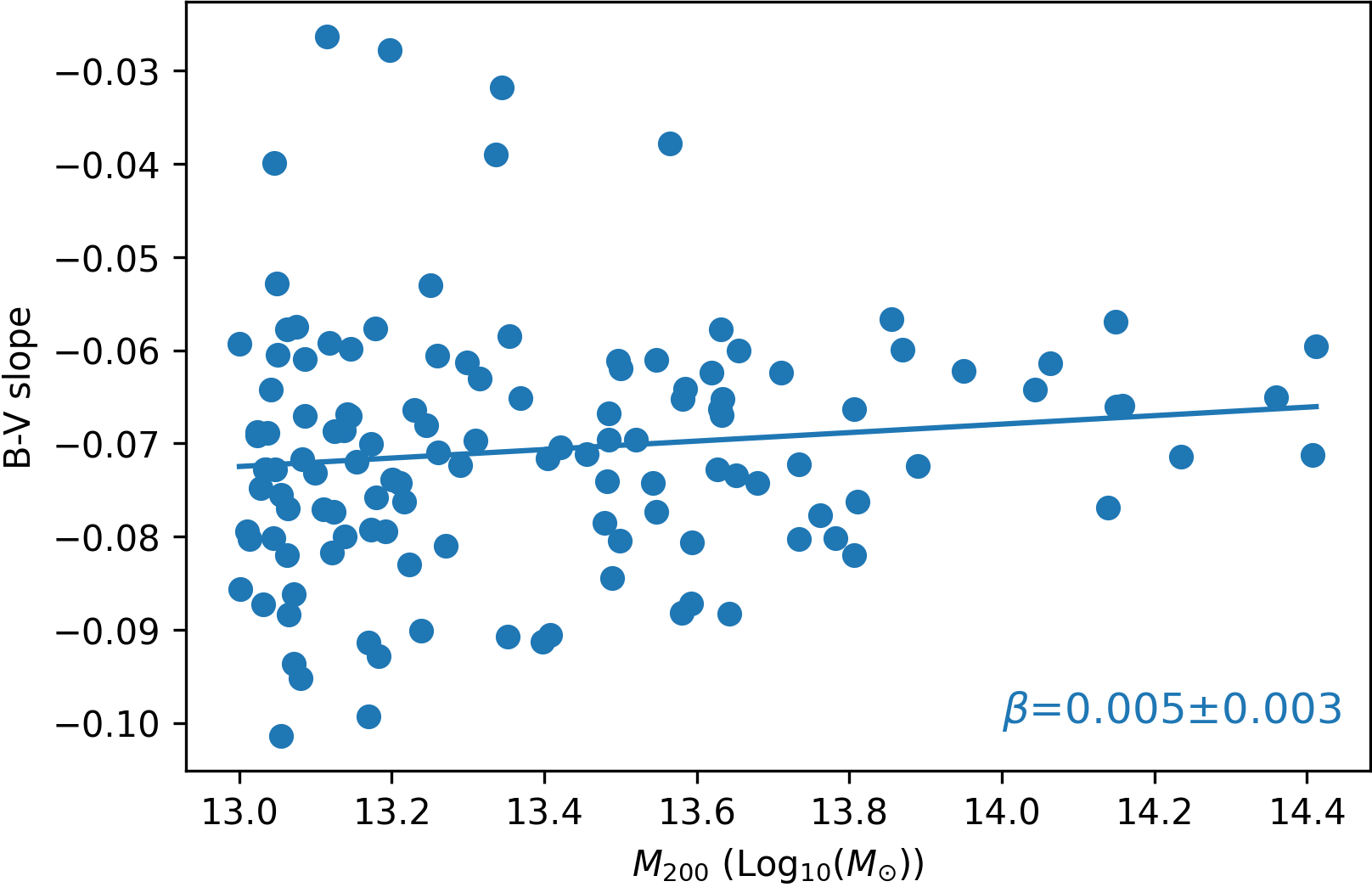}
    \caption{Top: Mean ICL and BCG colour, B-V. Middle: Mean B-V colours of the three different components of the BCG+ICL. Bottom: Cluster slopes of the colour, B-V, of the BCG+ICL plotted over the full cluster radius from the centre of the BCG out to $R_{200}$. The slopes of the linear fit to the data, $\beta$, are indicated in the bottom centre and right of the panel.}
    \label{fig:BV}
\end{figure}
The top panel of Fig.~\ref{fig:BV} plots the mean B-V colour (Vega magnitudes using an X filter from \citet{Buser1978}, where X=U, B, V)  of the ICL and BCG and shows that the ICL is also significantly bluer than the BCG. Both relations between the BCG colour and the ICL colour with cluster mass have small positive slopes, indicating that they become redder at higher masses. We note, however, that this trend is very weak. As with the metallicity plot, there is more scatter in the colours of the ICL. The middle panel of Fig.~\ref{fig:BV} shows the colour of the three different components that contribute to the BCG+ICL systems and also shows that the surviving galaxies component is bluest, followed by the in situ component and the component from completed mergers being redder, following the trend in the metallicity. The mergers component has a slight positive slope. The in situ component has a negative slope, but larger scatter than the mergers component, enough that the slope is consistent with a flat or positive trend. The surviving component is the only component with a clear trend, \hl{reflecting that galaxies are redder in more massive clusters \mbox{\citep{Budzynski_2013}}}.  
It also shows the greatest scatter of the three components, reflecting the strong individual variation in the formation histories of different clusters. The bottom panel of Fig.~\ref{fig:BV} plots the B-V gradient of different clusters and groups over the BCG+ICL region by the same method used in the bottom panel of Fig.~\ref{fig:Z}. 

As with the metallicity slope, the relation is universally negative, with no clusters showing any positive gradients. Unlike the metallicity plot, the relation shows a slight increase with mass, indicating that higher-mass clusters have less negative gradients, possibly due to increased stripping of redder galaxies at higher radii. The relation also shows greater scatter at lower masses, indicating greater variation in formation histories at low masses.

\subsection{Age}

In this section, we consider the ages of the ICL and BCG, as well as the ages of the three different components that make up the BCG+ICL. We note that age here refers to the stellar formation time of the particles that make up the stellar systems rather than their assembly times. A detailed consideration of how the ICL and the BCG form requires both, but we defer this to a later paper.

\begin{figure}

        \includegraphics[width=0.95\columnwidth]{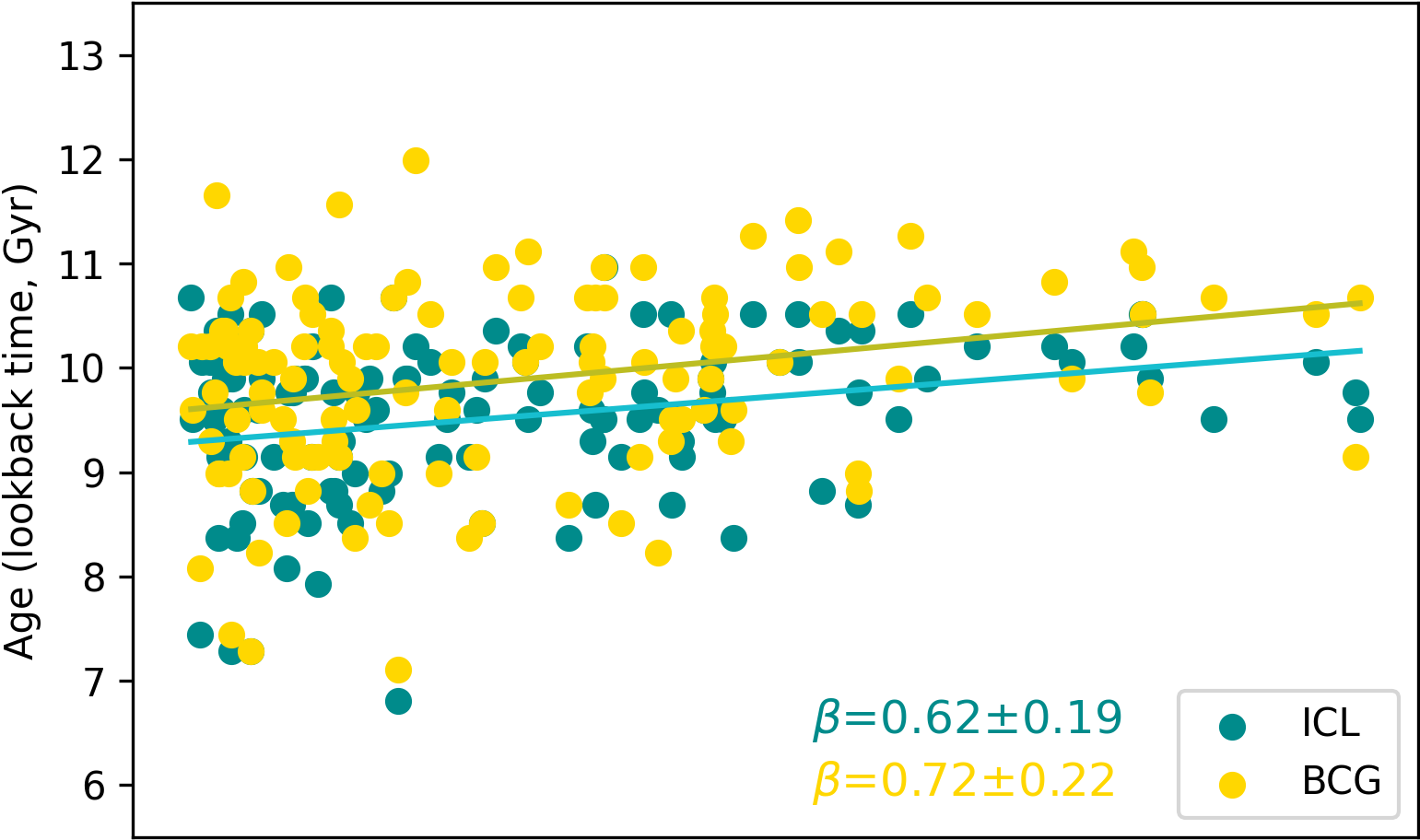}
    \includegraphics[width=0.95\columnwidth]{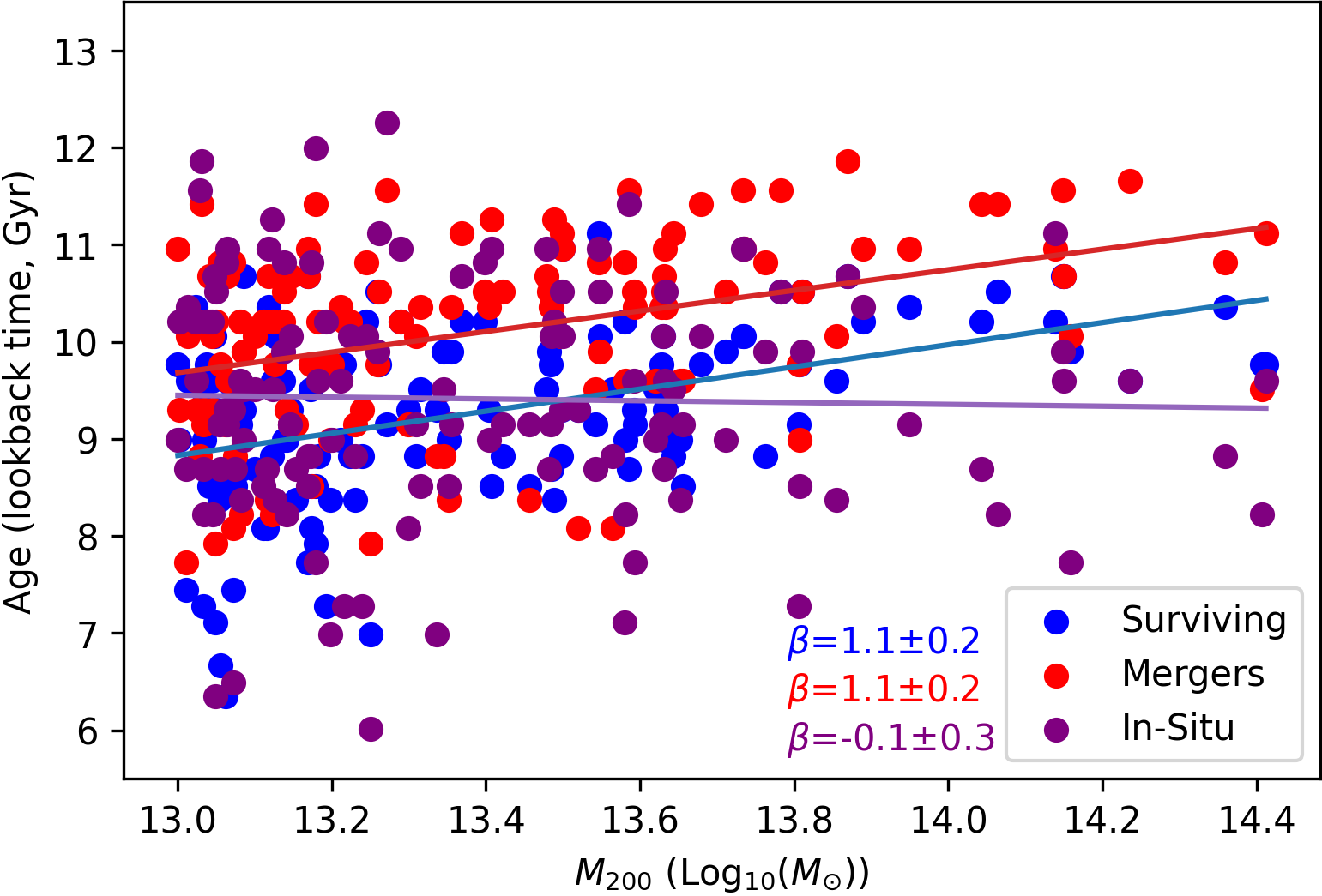}
    \caption{Top: Mean ICL and BCG ages (in Gyr). Bottom: Mean ages of the three different components of the BCG+ICL (in Gyr). The slopes of the linear fit to the data, $\beta$, are indicated in the bottom right of the panel.}
    \label{fig:snapage}
\end{figure}

The top panel of Fig.~\ref{fig:snapage} shows the mean age of the stellar particles, in lookback time, of the ICL and the BCG. The ICL is found to be slightly younger than the BCG, although both distributions have a lot of scatter, and the differences are within the range of scatter. Both the ICL and the BCG display a trend where the mean age increases with cluster mass; however, at a mass of approximately $M_{200}$~$=$~8~$\times$~10\textsuperscript{13}~\(\textup{M}_\odot\), the trend appears to reverse and the ages of both the ICL and BCG appear to decline with mass (although the difference is still within the range of scatter). The bottom panel of Fig.~\ref{fig:snapage} plots the mean ages by component. The mean age-cluster mass relation of the in situ component is largely flat. The relation exhibits considerable scatter, particularly at lower masses. The completed mergers component 
has a positive slope, where mean age increases with cluster mass and is the oldest of the three components. The surviving component has a slope consistent with the merging component, while being younger. At lower masses, the surviving is the youngest of the three components; at the highest masses, it is intermediate between the merging component and the in situ component. 

\section{ICL and BCG assembly histories}
\label{section:ICLprogs}

In this section, we investigate the properties of the progenitor galaxies that formed the ICL and the BCG.

\subsection{ICL and BCG progenitor masses}

\begin{figure}

        \includegraphics[width=0.95\columnwidth]{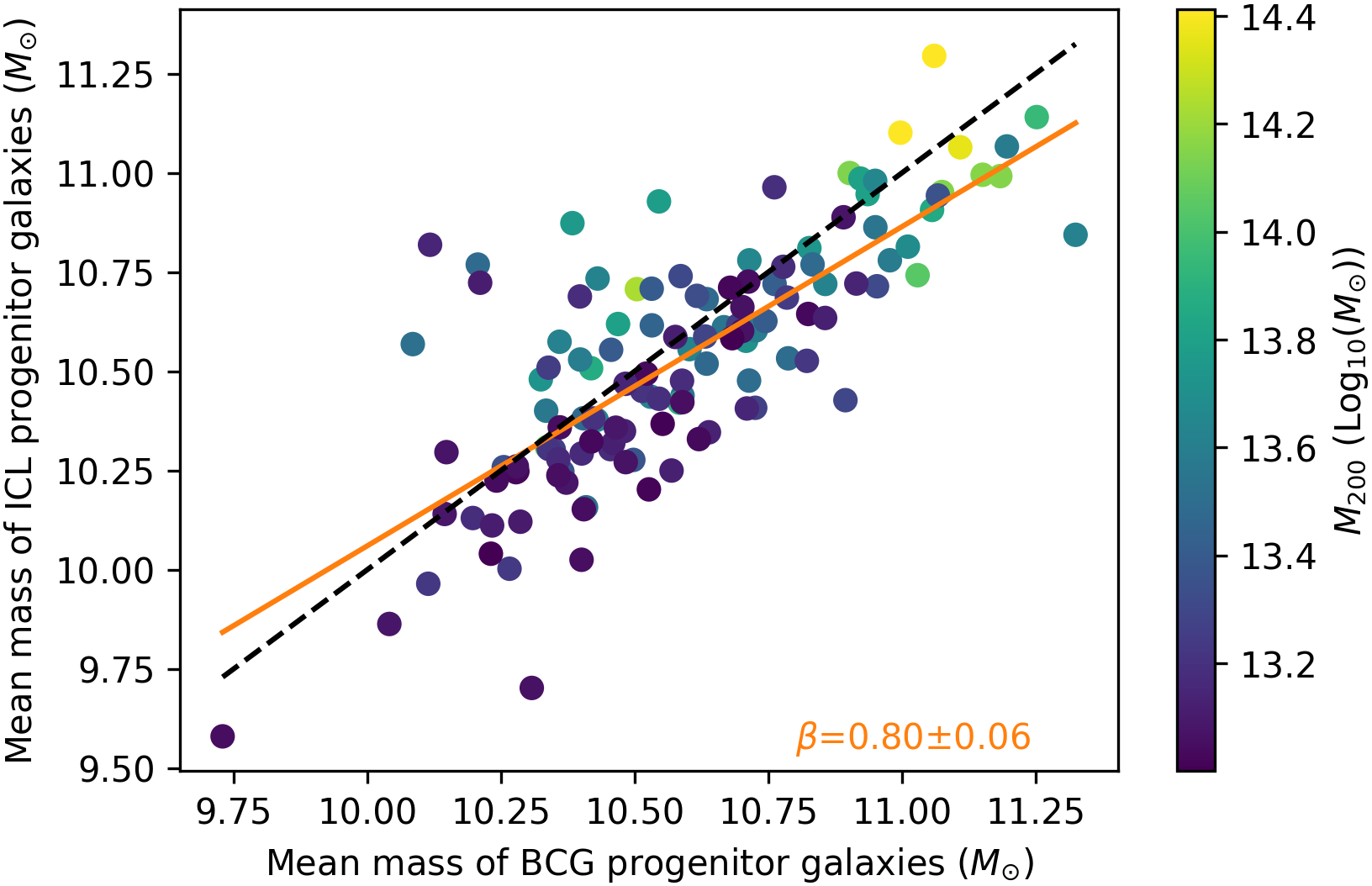}
    \caption{Mean mass of progenitor galaxies that formed the BCG and ICL, coloured by the cluster mass. The orange line is the line of best fit. The black dashed line is the one-to-one ratio. The slope of the linear fit to the data, $\beta$, is indicated in the bottom right of the panel. 
    }
    \label{fig:progmeanmass}
\end{figure}

\hl{The mean mass of progenitor galaxies in the ICL and BCG is calculated by taking the mean mass of all progenitors, weighed by the number of particles a progenitor contributes. Note that we do not impose a minimum mass threshold on our progenitor galaxies, but galaxies with mass \mbox{M\textsubscript{*}~<~1~$\times$~10\textsuperscript{7}~\(\textup{M}_\odot\)} only make up approximately 0.1 per cent of the sample}. Figure~\ref{fig:progmeanmass} plots the mean mass of the BCG and the ICL progenitor galaxies against one another and shows that the mean mass of ICL progenitor galaxies tracks slightly behind the mean mass of BCG progenitors for a given BCG progenitor mass. The mean mass for both increases with cluster mass; however, there is a lot of scatter in the relation, which suggests that there is a lot of variation in the original galaxies that make up the ICL and the BCG, even at a given cluster mass. 

Also notable is the fact that the colour bar shows that \hl{if we fix the mean mass for the BCG progenitor galaxies, and increase the total cluster mass the mean mass of the ICL progenitor galaxies increases. This is likely because more massive clusters have a satellite mass function shifted towards higher masses. The population of satellites susceptible to stripping is more massive than in lower-mass clusters. Because the ICL is dominated by the stripping of intermediate- and high-mass galaxies rather than dwarfs, the mean mass of its progenitors increases even if the mean progenitor mass of the BCG remains unchanged. Additionally, in the most massive clusters, the high velocity dispersion tends to lengthen dynamical friction timescales. As a consequence, many massive satellites that would merge with the BCG in lower-mass halos are not captured by z=0. However, these same galaxies do undergo efficient tidal stripping within the deep cluster potential. Thus, even if the typical mass of BCG progenitors remains fixed, the ICL is increasingly fed by more massive satellites, causing the mean progenitor mass of the ICL to rise with cluster mass.}

The mean fraction of ICL mass contributed by low-mass progenitors (M\textsubscript{*}~<~1~$\times$~10\textsuperscript{10}~\(\textup{M}_\odot\)) is $34.6 \pm 14.9$ per cent, and the mean fraction of  ICL mass contributed by high-mass progenitors (M\textsubscript{*}~>~1~$\times$~10\textsuperscript{10}~\(\textup{M}_\odot\)) is $65.4 \pm 14.9$ per cent. \hl{This is consistent with previous simulations, which also find that the ICL is typically dominated by a few progenitors with masses of approximately \mbox{M\textsubscript{*}~>~1~$\times$~10\textsuperscript{10}~\(\textup{M}_\odot\))}.} \citep[e.g.][]{contini2014, Contini_2020, Cooper20150, Harris_2017, Ahvazi2024}. \hl{This is also consistent with observational studies which have used colours and profiles to infer that the ICL forms primarily from the disruption of intermediate-mass galaxies \mbox{\citep[e.g.][]{Montes_2014, DeMaio2018}.} }

To compare the properties of clusters with ICL dominated by low- and high-mass progenitors, we select clusters outside of one standard deviation of the mean. We selected 16 clusters that are dominated by low-mass galaxies, with the fraction of low-mass galaxies in the ICL being greater than 49.5 per cent, and 16 clusters that are dominated by high-mass galaxies, with the fraction of low-mass galaxies in the ICL making up less than 19.0 per cent. 
A high fraction of mass contributed by low-mass galaxies in the ICL means the cluster is likely to be in an intermediate relaxation state. Next, 93.8 per cent (15/16) of the selection of clusters dominated by low-mass galaxies were in an intermediate (\hl{transitioning between disturbed and relaxed}) dynamical state. Only 6.3 per cent (1/16) were relaxed and none were disturbed. A large fraction of mass contributed by high-mass galaxies to the ICL means the cluster is more likely to be disturbed, with 31.3 per cent (5/16) being disturbed and 25.0 per cent (4/16) relaxed. They are also less likely to be in an intermediate state, making up a fraction  of 44.0 per cent (7/16). This suggests that highly evolved (relaxed) clusters and clusters in an active state of disruption (disturbed) are more likely to have a significant contribution from high-mass galaxies in the ICL. In contrast, clusters in an intermediate state of relaxation lack accretion of high-mass galaxies into the ICL. \hl{This is likely because past or ongoing disruptions from massive galaxies characterise clusters in a relaxed or disturbed state. \mbox{\citet{Brown2024}} find that a major accretion event (which disturbs a cluster) temporarily forms a large amount of ICL. However, in a longer time frame, the ICL is dominated by intermediate-mass contributors \mbox{\citep[e.g.][]{Contreras_Santos_2024}.}} Clusters with ICL dominated by low-mass galaxies have a higher mean concentration of $6.6 \pm 1.5$, and lower in clusters where the ICL is composed of high-mass galaxies $5.2 \pm 1.6$. Although the differences are within 1 sigma uncertainty, this could be a result of the \hl{disruption} of low-mass galaxies being more efficient in more concentrated clusters. \citet{contini2014, Contini_2020} \hl{find that more highly concentrated clusters have a higher proportion of material contributed from low- and intermediate-mass progenitors}. High-mass galaxies in the ICL indicate a more recent major merger, although the differences are within 1 sigma uncertainty. \hl{Clusters dominated by high-mass galaxies, on average, experienced a major merger} $5.5 \pm 2.7$ Gyr ago versus \hl{a major merger} $8.1 \pm 2.6$ Gyr ago \hl{for clusters dominated by low-mass galaxies}.

A larger fraction of mass contributed by low-mass galaxies in the ICL indicates that the ICL is more metal-poor, $-0.55 \pm 0.07$ dex for the low-mass fraction and $-0.44 \pm 0.06$ dex for the high-mass fraction. This is expected from the mass-metallicity relationship, where more massive galaxies are more metal-rich \citep{Ma_2015}. 
The metallicity slope was also sharper for clusters with a higher percentage of low-mass galaxies in the ICL, with a mean slope of $-0.41 \pm 0.04$. In contrast, clusters with higher-mass galaxies in the ICL had a slope of $-0.34 \pm 0.04$. This would indicate that a lower ICL metallicity does not necessarily correspond with a lower BCG metallicity, possibly because the BCG has a much larger proportion of in situ star formation, which will be metal-rich in the massive BCG galaxies.

\subsection{Shared progenitors}
In this section we analyse the shared formation history of the ICL and the BCG by considering shared progenitor galaxies. This is done by determining whether progenitor galaxies that contributed to the growth of the ICL or the BCG also contributed to the growth of the BCG or the ICL, respectively. To determine shared accretion histories for each star particle found in the ICL, we first determine its progenitor galaxy. We then check whether this progenitor galaxy also contributed any star particles to the BCG. This is repeated for every star particle in the ICL, giving us the percentage of star particles in the ICL that share progenitor galaxies with the BCG. The same process is then repeated for particles in the BCG. We then have a final percentage of particles in the ICL that share progenitors with the BCG and vice versa. Note that this analysis does not consider the relative contribution each of these progenitor galaxies makes to the ICL and the BCG, only whether they contribute to both. To determine the relative contribution, an analysis of the most significant progenitors is necessary. {Significant progenitors are those that contributed to the build-up of 90 per cent of the material of either the ICL or the BCG. The most significant progenitor is the single progenitor galaxy that contributed the most material to either the ICL or the BCG}. We address in part the influence of the ICL and BCG sharing their most significant progenitor in the section above and Sect. \ref{section:corrcomptop}, but we defer a more detailed analysis to a future paper.

\begin{figure}

    \includegraphics[width=0.95\columnwidth]{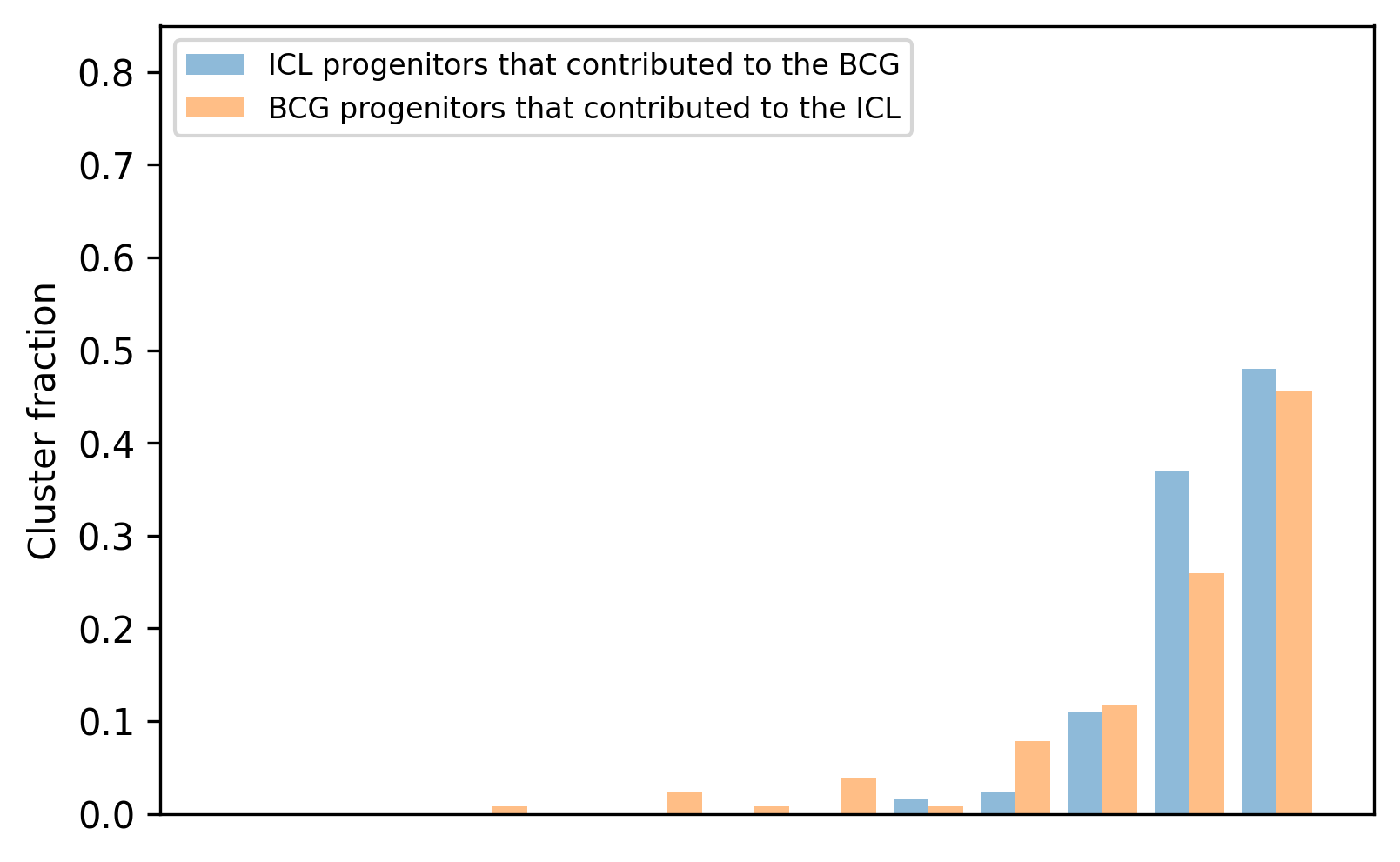}
        \includegraphics[width=0.95\columnwidth]{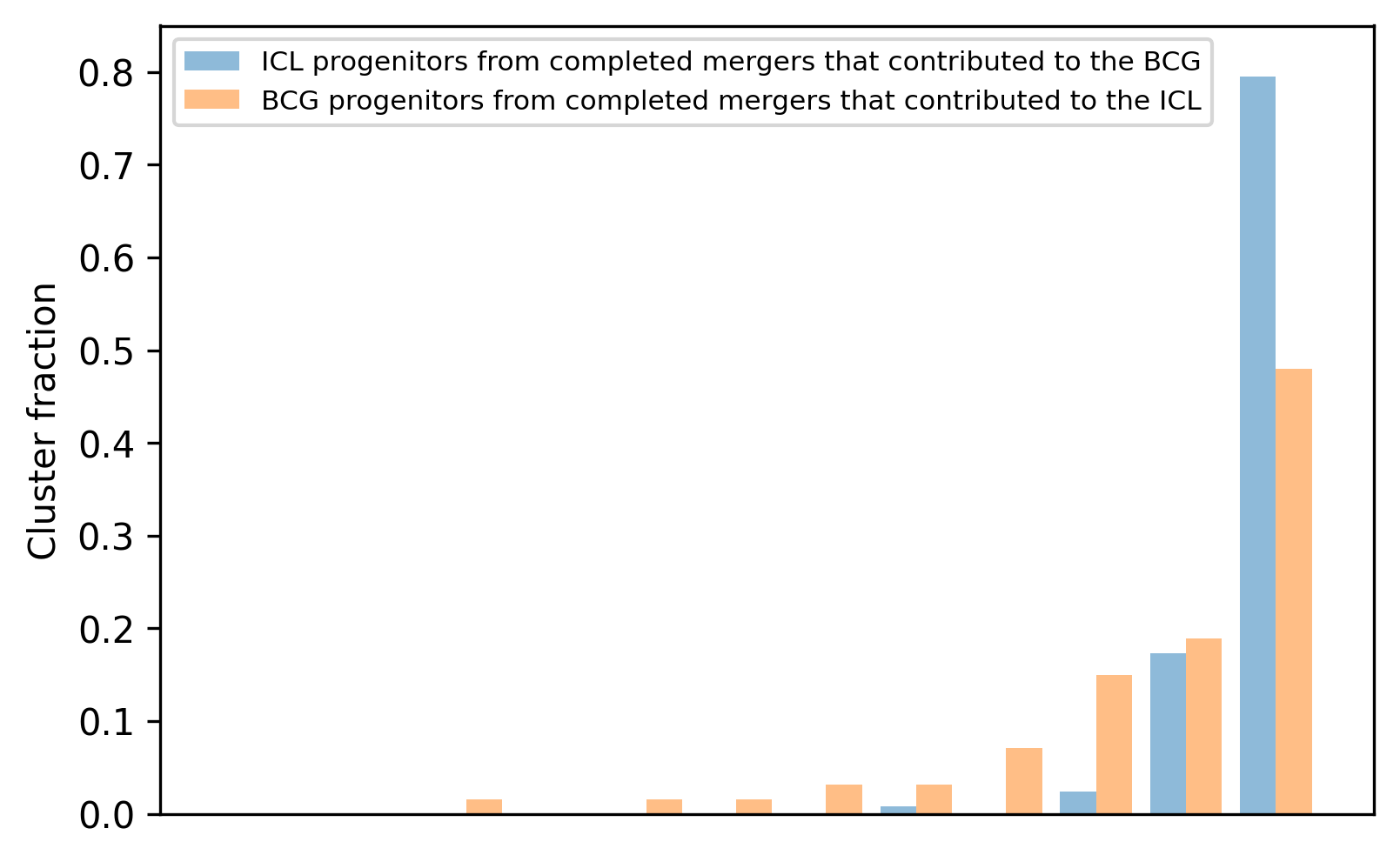}
 \includegraphics[width=0.95\columnwidth]{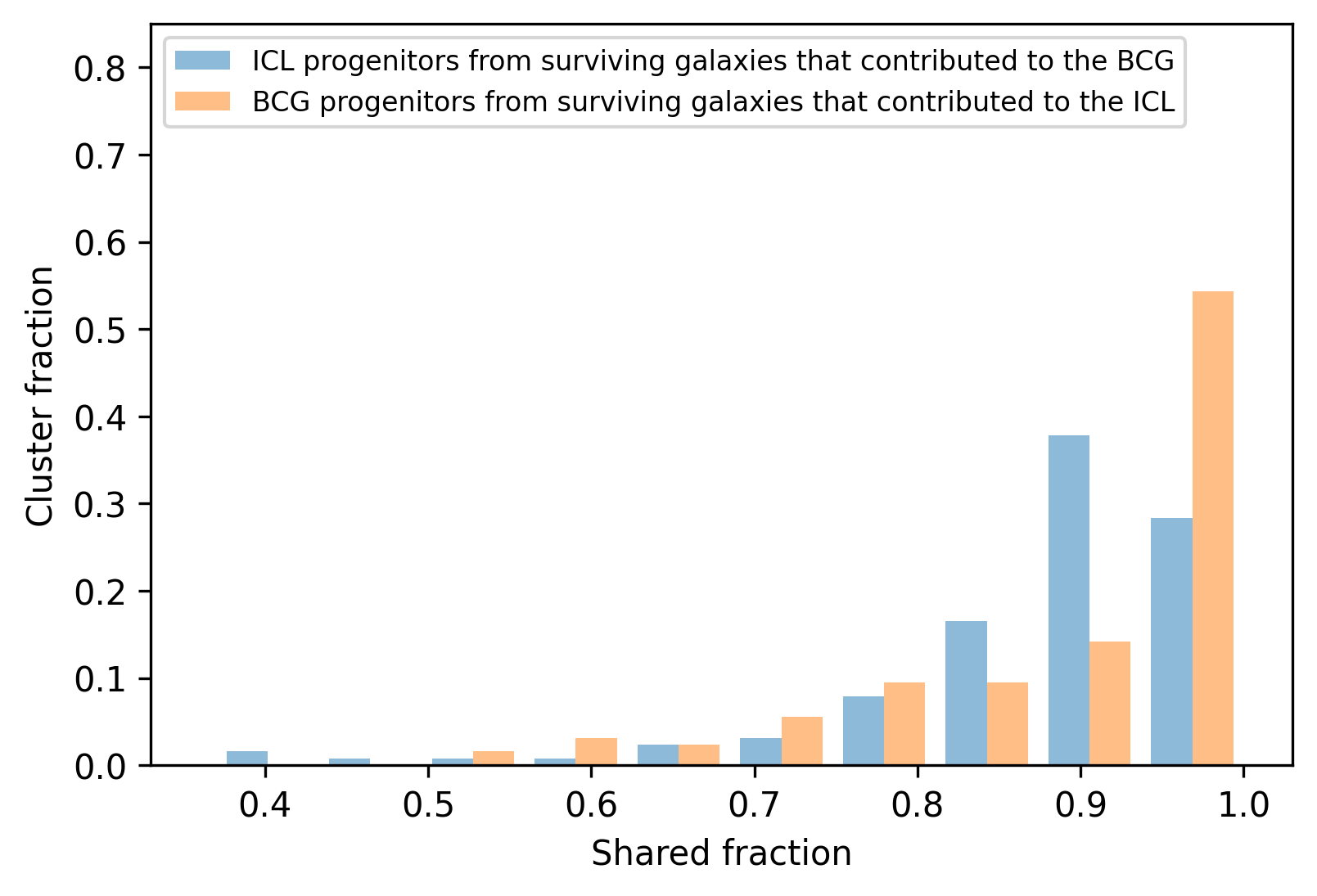}
    \caption{Fraction of \hl{particles that share progenitor galaxies} between the ICL (blue) and BCG (orange). \hl{A high shared fraction means that a large number of particles in the ICL originate in progenitor galaxies that also contribute to the BCG and vice versa}. The top panel shows the overall shared fraction, while the bottom two panels divide the shared fraction of progenitors from completed mergers (middle) and progenitors from surviving galaxies (bottom).}
    \label{fig:ICLBCGprogs}
\end{figure}

The top panel of Fig.~\ref{fig:ICLBCGprogs} shows the fraction of \hl{particles of the ICL that share progenitors with the BCG and particles of the BCG that share progenitors with the ICL, respectively}. \hl{A high shared fraction means that a large number of particles in the ICL originate in progenitor galaxies that also contribute to the BCG and vice versa.} This plot shows that in the majority of cases, the BCG and the ICL have shared accretion histories. The overall mean percentage of shared progenitors is 93 per cent for the ICL and 91 per cent for the BCG. Overall, there are a larger number of galaxies that contribute to the BCG growth but not ICL growth, although the number of galaxies found in the BCG that are also found in the ICL is still very high. This panel also shows a long tail of clusters where the shared progenitors of the ex situ component of the BCG are lower. The lowest fraction of shared progenitors in the ICL is 0.77, but the lowest fraction of shared progenitors for the BCG is 0.50, indicating that overall, the galaxies that contribute to the growth of the ICL will also contribute to the growth of the BCG, but in some cases, ex situ BCG growth can occur independent of ICL growth. For 85 per cent of the clusters (108/127), over 90 per cent of the accreted ICL mass fraction is contributed by progenitors that also contributed to the growth of the BCG. For the BCG, the number drops to 72 per cent (91/127). It is worth noting that even though a large number of clusters shared a significant number of ICL and BCG progenitors, the properties of the BCG are likely not a good indicator of the ICL formation history due to the large contribution of in situ material to the BCG, especially in low-mass structures.
The lower two panels show the distinct contributions of material from accreted galaxies, completed mergers (middle) and surviving progenitors (bottom). The middle plot shows that completed mergers contributing to the growth of the ICL are also likely to be found in the BCG. However, there are a number of clusters where the BCG has experienced mergers with progenitor galaxies that are not found in the ICL. The lowest figure panel shows the shared contribution of surviving galaxies to the ICL and BCG and shows that, in this case, progenitor galaxies can contribute to ICL growth but not BCG growth, which suggests that these clusters may be experiencing stripping in the outskirts of the cluster that has not contributed to the growth of the BCG. Galaxies that contribute to the growth of either the ICL or the BCG but not the other are, on average, less massive than the overall mean masses of galaxies simultaneously contributing to both. 

To compare the properties of clusters with high and low similarities in accretion history, we first select 17 clusters where the ICL shared less than 89.5 per cent of progenitors with the BCG and 14 clusters where the BCG shared less than 82 per cent of progenitors with the ICL. Secondly, we chose 17 clusters where the ICL shared more than 96.9 per cent with the BCG and 14 clusters where the BCG shared more than 98.8 per cent of progenitors with the ICL. 

Clusters with ICL composed of progenitor galaxies that were shared with the BCG were more likely to be in a relaxed state than clusters where the ICL did not share galaxies with the BCG. For the high shared selection, 12 per cent (2/17) of the clusters were disturbed, 41 per cent (7/17) were relaxed, and 47 per cent (8/17) were intermediate, while for the low shared selection, 35 per cent (6/17) of the clusters were disturbed, 24 per cent (4/17) were relaxed, and 41 per cent (7/17) were intermediate. Clusters with BCG composed of progenitor galaxies that were shared with the ICL were more likely to be disturbed than clusters where the BCG did not share progenitor galaxies with the ICL. For the high shared selection, 36 per cent (5/14) were disturbed, while for the low shared selection, 21 per cent (3/14) were disturbed. Clusters with a higher shared fraction were more likely to be relaxed than those with a low fraction, 43 per cent (6/14) versus 14 per cent (2/14), and less likely to be in an intermediate state, 21 per cent (3/14) versus 64 per cent (9/14). This indicates that in highly evolved, relaxed clusters, it is more likely for the ICL and BCG to have similar accretion histories. In contrast, younger clusters or those that have not undergone significant accretion events in the past are more likely to have distinct ICL and BCG populations. Clusters where the ICL shared progenitor galaxies with the BCG were unlikely to be disturbed, but clusters where the BCG shared galaxies with the ICL were likely to be disturbed, a likely indicator of the greater influence major mergers have on the BCG than the ICL. Clusters with ICL composed of progenitor galaxies that were shared with the BCG were more highly concentrated than those with a low shared fraction, but within 1 sigma of uncertainty $6.4 \pm 1.3$ versus $4.8 \pm 1.8$.  This is possibly because in more highly concentrated clusters, accretion is more efficient, and galaxies that experience accretion into the ICL will more rapidly merge with the BCG. \hl{Clusters with BCG composed of progenitor galaxies that were shared with the ICL were not more or less concentrated}.
\hl{The ICL and the BCG grow by the same processes, where material from accreting galaxies is initially stripped into the ICL, and the progenitor galaxies subsequently merge with the BCG. Therefore, the processes that build ICL and BCG overlap but operate on different timescales and radii} \citep{contini2014, Contini_2018, Harris_2017}.

\hl{In summary, the ICL and the BCG generally share progenitors. In both, there is a high likelihood that if particles from a progenitor galaxy are found in the ICL they will also be found in the BCG and vice versa. The lowest shared fractions were found for completed progenitor mergers in the BCG and surviving progenitors in the ICL. This is likely from progenitor galaxies that did not experience any merger events until they reached the central BCG (thus not contributing to the ICL), and galaxies that have just begun to experience stripping in the ICL, and have not yet reached the BCG.}

\subsection{Locations of material from high- and low-mass progenitor galaxies}

\begin{figure}

    \raggedleft
        \includegraphics[width=0.95\columnwidth]{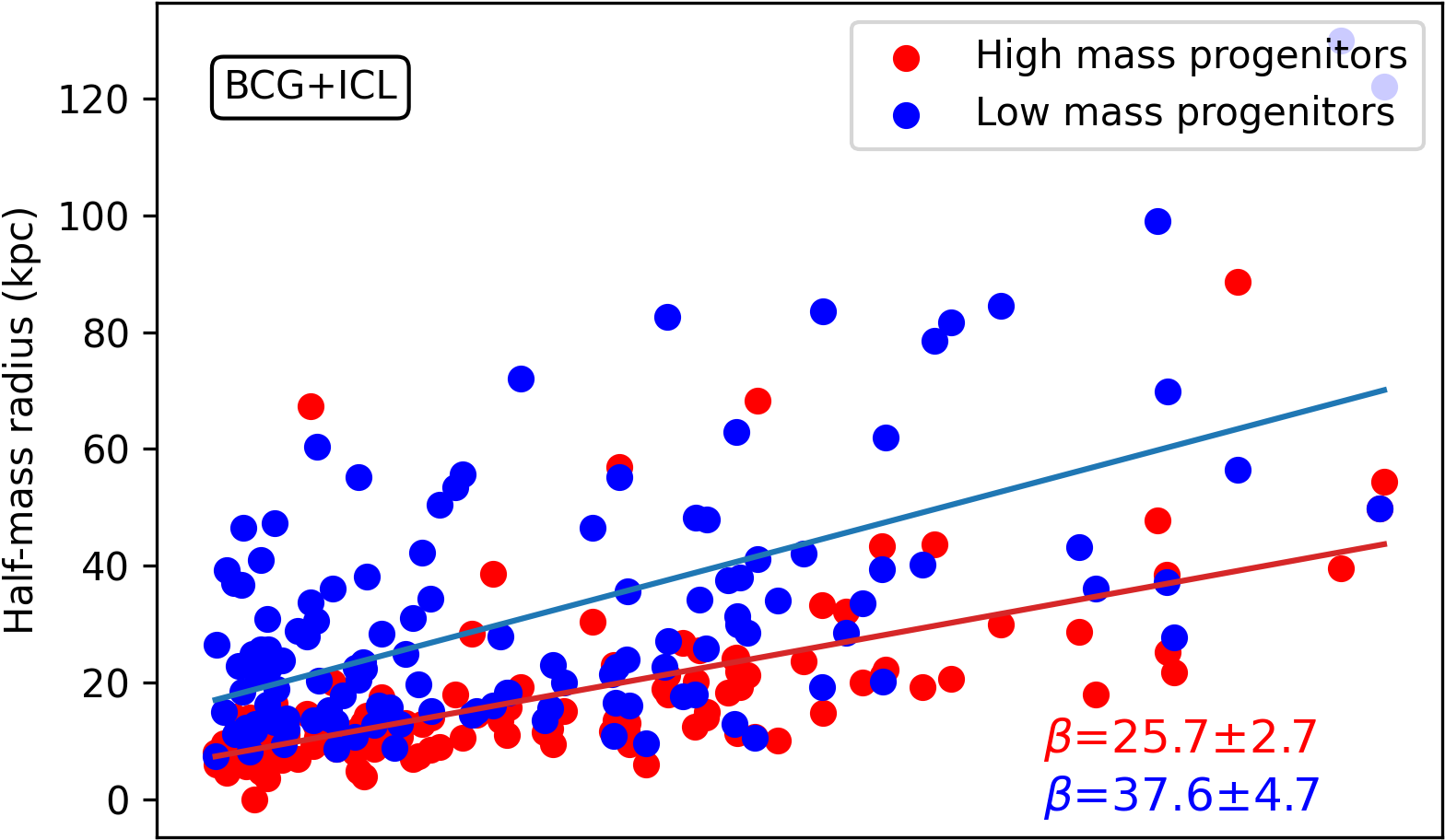}
        \includegraphics[width=0.934\columnwidth]{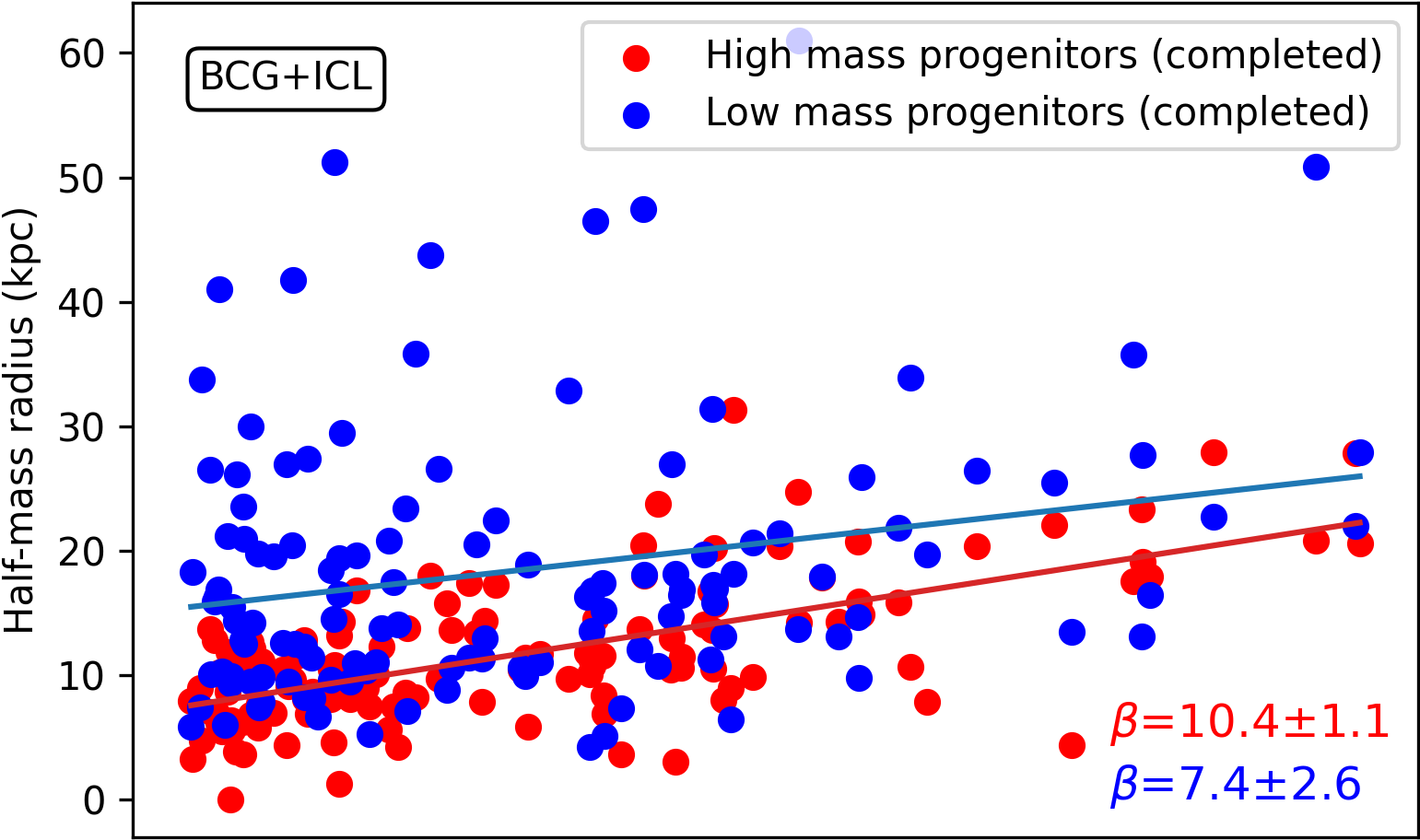}
        \includegraphics[width=0.95\columnwidth]{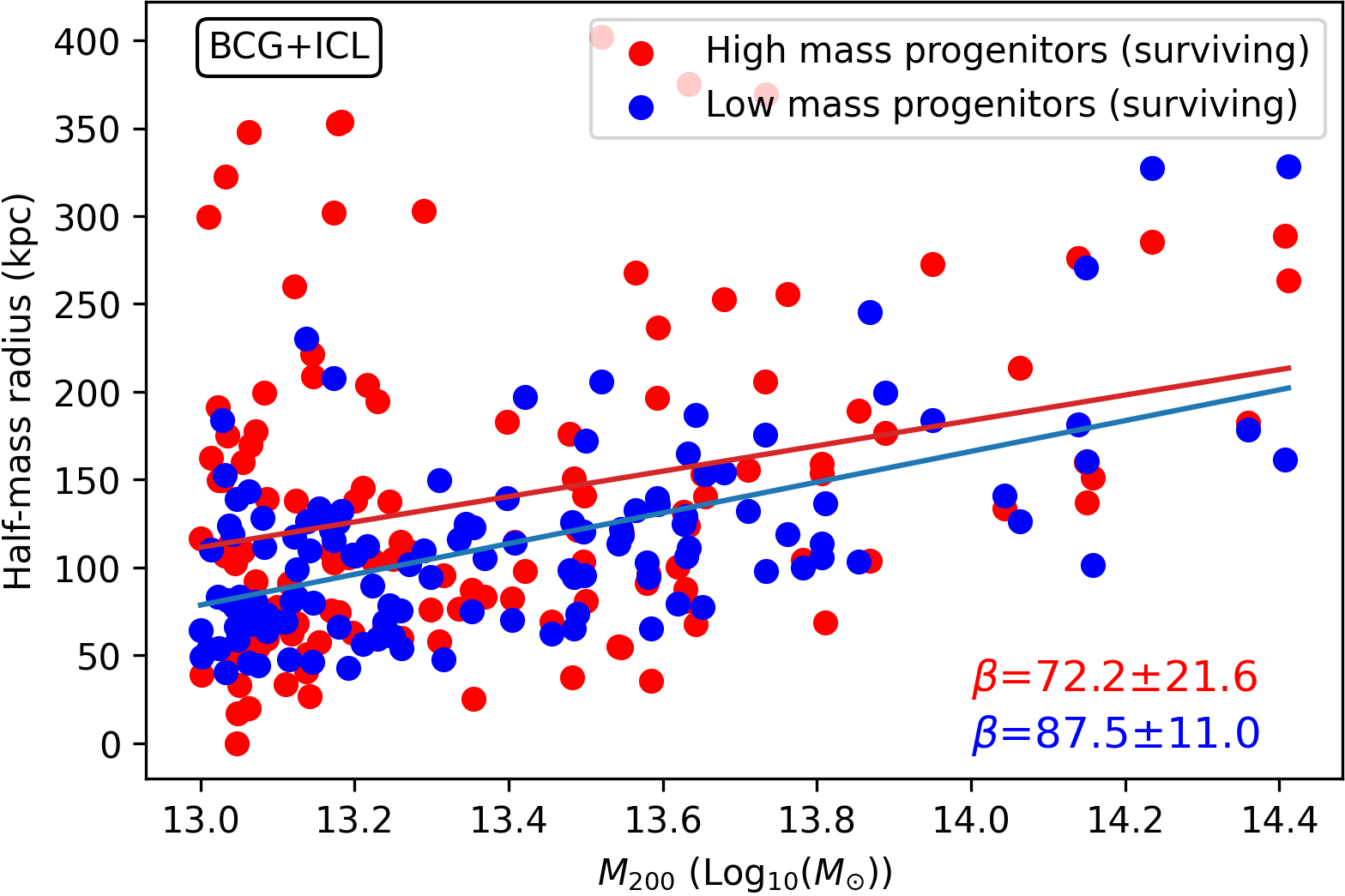}
    \caption{Top: HMRs of material accreted from low (M\textsubscript{*}~<~1~$\times$~10\textsuperscript{10}~\(\textup{M}_\odot\)) and high (M\textsubscript{*}~>~1~$\times$~10\textsuperscript{10}~\(\textup{M}_\odot\)) mass progenitor galaxies that formed the BCG+ICL. 
    Middle: HMRs of material accreted from completed mergers. 
    Bottom: HMRs of material accreted from surviving galaxies. The slopes of the linear fit to the data, $\beta$, are indicated in the bottom right of the panel.
    }
    \label{fig:projdist}
\end{figure}

Figure~\ref{fig:projdist} plots the HMRs of the stellar mass distribution from low (M\textsubscript{*}~<~1~$\times$~10\textsuperscript{10}~\(\textup{M}_\odot\)) and high-mass (M\textsubscript{*}~>~1~$\times$~10\textsuperscript{10}~\(\textup{M}_\odot\)) progenitor galaxies that contributed to the growth of the BCG+ICL. The top panel does not separate the different formation channels. The middle panel shows progenitor galaxies that have completed mergers, and the bottom panel shows material stripped from surviving galaxies. The contribution from high-mass galaxies is more centrally concentrated than that from low-mass galaxies, suggesting that accretion from high-mass galaxies dominates the central areas of the BCG. In contrast, outer areas are largely accreted from lower-mass galaxies. However, the lower two panels show that there is a clear difference in the concentration of particles from low- and high-mass galaxies, depending on their origin, and in some cases, the HMR of the contribution from surviving galaxies is almost an order of magnitude larger than that from completed mergers. The central panel shows that material from high-mass galaxies that have completed the merger process is more centrally concentrated than material from low-mass galaxies. Conversely, the bottom panel shows that material from low-mass progenitor galaxies that have undergone stripping, but not been completely disrupted, is more centrally concentrated than material from high-mass galaxies, likely because high-mass galaxies experience more dynamical friction and will start being stripped at larger radii. Note that in both of the two lower panels, the $r_h-M_{200}$ distributions of both high and low-mass progenitors are consistent with each other.

This suggests a transition where the inner areas of the BCG are dominated by stars from high-mass galaxy mergers, the outer areas where the BCG and ICL begin to overlap and are dominated by stars from dwarf galaxy mergers; and finally, the outskirts of the ICL are largely composed of stars stripped from high-mass galaxies. This likely occurs because high-mass galaxies will only be completely disrupted in the central parts of the cluster, instead experiencing tidal stripping but surviving at higher radii. In contrast, low-mass galaxies that experience tidal stripping are more likely to be completely disrupted rather than experience an extended stripping period before disruption \hl{\mbox{\citep[e.g.][]{Annunziatella_2016, Bah__2019}}}. The middle panel shows a large scatter in the \hl{HMR} of accreted low-mass galaxies that have completed mergers, with many being \hl{disrupted} at high radii. In contrast, the bottom panel shows a larger scatter in the HMR of high-mass progenitors, with many experiencing stripping at high radii. \citet{Cooper20150, Harris_2017} and \citet{Ahvazi2024} \hl{all found that massive progenitors tend to be found at lower radii, in the outer BCG envelope or inner ICL, while lower-mass progenitors build up the extended ICL, consistent with our results that material from high-mass galaxies is generally more concentrated.}

\subsection{Correlation between components and significant progenitors}
 \label{section:corrcomptop}
In this section we discuss correlations between the three different components that make up the ICL and BCG, and progenitor galaxies that contribute to their growth. Note that the in situ component is not directly affected by accreted progenitor galaxies. However, we also consider the significance of this component as a proxy for clusters that have experienced high or low amounts of accretion, as the other two components are both part of the same process of accretion.

\subsubsection{Surviving component} 

In this section we select clusters with a high proportion of material from surviving galaxies, using the same selection criterion as in Sect. \ref{section:compcorr}, to compare with progenitor properties. The mean mass of progenitor galaxies that contribute to the ICL is found to be larger in clusters with a higher fraction of material coming from surviving galaxies, although consistent within 1 sigma uncertainty.  
The upper selection has a mean progenitor mass of $10.8 \pm 0.2$ \(\textup{M}_\odot\), and the lower selection $10.5 \pm 0.3$ \(\textup{M}_\odot\). \hl{This is likely because stellar stripping is a larger contributor in higher-mass clusters, which also have higher-mass satellites. Thus, systems with a high fraction of stripping are more likely to be host to more massive galaxies. } 
A large surviving component in the ICL decreased the likelihood that the BCG and the ICL {had the same most significant progenitor. Significant progenitors are those that contributed 90 per cent of the material of the ICL and BCG. The most significant progenitor is the single progenitor galaxy that contributed the most material to the ICL and BCG}. For the upper selection of clusters, none share their most significant progenitor with the BCG, 
By contrast, most clusters with low percentages of material stripped from surviving galaxies in the ICL share the most significant progenitor between the ICL and the BCG, at 63 per cent (17/27). 
For the clusters with a higher fraction of surviving material in the ICL, $24 \pm 9$ per cent of the material was from low-mass progenitors and $76 \pm 9$ per cent from high-mass progenitors. In comparison, the clusters with a lower fraction of surviving material in the ICL have $36 \pm 16$ per cent from low-mass progenitors and $64 \pm 16$ per cent from high-mass progenitor galaxies. 
\hl{On average, the contribution of the most significant progenitor to the accreted mass of the ICL is \mbox{$0.27 \pm 0.12$} per cent. This was found to be lower for clusters with a higher fraction of surviving material in the ICL, although within uncertainties. The fraction with higher surviving ICL had \mbox{$25 \pm 19$} per cent material accreted from a single galaxy, while the lower fraction had \mbox{$35 \pm 14$} per cent.}

\subsubsection{Completed mergers' component}

Unlike the surviving component, no significant difference was found in mean progenitor mass 
for the completed mergers component. ICL with a high fraction of the completed mergers component are significantly more likely to share the most significant progenitor with the BCG than the lower fraction sample. 54 per cent (13/24) of the upper ICL sample 
share the same most significant progenitor, while only 1/24 (4 per cent) of the ICL with a low mass fraction coming from completed mergers 
share the most significant progenitor. 
The mass fraction coming from low- and high-mass progenitor galaxies is found to be consistent for both the high and low completed mergers fraction. No significant difference is found in the mean mass of the ICL progenitor galaxies. The fraction the most \hl{significant progenitor} made up of the ICL is found to be higher for clusters where the completed mergers component is higher, although within uncertainties. The fraction one galaxy made up of the ICL is $0.37 \pm 0.14$ for the upper ICL sample and  $0.24 \pm 0.10$ for the lower ICL sample. 

\subsubsection{In situ component}
The in situ component is not strongly affected by galaxy accretion, but it can be used as an indirect indicator of ex situ component significance. This is because the fraction of ex situ mass on the given structure will also depend on the in situ component.
Clusters with a higher in situ fraction in the ICL are likelier to share the most significant progenitor with the BCG than the lower fraction sample. 47 per cent (9/19) of the clusters with a higher in situ fraction in the ICL 
share the same most significant progenitor, while only 16 per cent (3/19) of the clusters with a lower in situ fraction in the ICL share the \hl{most significant progenitor}. 
Clusters with a higher in situ fraction in the ICL are likelier to have a high fraction of low-mass progenitor galaxies in the ICL. The clusters with a higher in situ fraction in the ICL have a low mass fraction of $51 \pm 26$ per cent, and the lower clusters $22 \pm 7$ per cent.
Correspondingly, they are less likely to have a high fraction of high-mass progenitor galaxies. The clusters with a higher in situ fraction in the ICL have a low mass fraction of $49 \pm 16$ per cent, and the lower clusters $78 \pm 7$ per cent. 
Additionally, there is a large difference in the mean mass of the ICL progenitor galaxies depending on the in situ fraction. The clusters with a higher in situ fraction in the ICL have an average progenitor mass of $10.2 \pm 0.2$ \(\textup{M}_\odot\), and the lower clusters $10.9 \pm 0.2$ \(\textup{M}_\odot\). 

It may seem surprising that the in situ component has a larger impact on progenitor galaxy mass than the stripped and merged components. However, it is worth noting that although these components are differentiated in our study, they both ultimately stem from the same processes. While mergers can trigger in situ stellar formation, it also grows from the formation of the BCG alone. As such, when the ICL has a large percentage of in situ formation, it indicates that less galaxy stripping and disruption are occurring in the cluster, which is likely due to these clusters hosting fewer galaxies and satellite galaxies with lower masses, and thus, fewer mergers and major mergers are occurring. Distinct from the stripped and merged fraction, the in situ component has no impact on the fraction  the most \hl{significant progenitor} makes up of the ICL and BCG+ICL. 

\section{Conclusions}
\label{section:conc}
In this work, we use the hydrodynamic IllustrisTNG-100 simulation to investigate the coevolution of the ICL and BCG, depending on formation history and properties. We selected 127 clusters and groups with masses >10\textsuperscript{13}~\(\textup{M}_\odot\), then separated the ICL from the BCG with a surface-brightness cut at the Holmberg radius of 26.5 mag/arcsec\textsuperscript{-2}. We summarise our results below. 

\subsection{ICL mass fraction}
\begin{enumerate}
    \item The ICL mass \hl{fraction} increases weakly with cluster mass, with large variation at a given mass. No clear correlation is found between ICL mass fraction and cluster relaxation.
    \item The ICL mass \hl{fraction} declines weakly with concentration. At a given concentration, more massive clusters have higher ICL mass fractions.
    \item Clusters with a more recent major merger contain lower ICL mass fractions.
\end{enumerate}

\subsection{ICL and BCG components}
\begin{enumerate}
    \item The surviving galaxies component of the BCG+ICL system is the most extended. The in situ component is the most central. The completed merger component lies in between and overlaps with the in situ component.
    \item The completed mergers component is the largest component of the BCG+ICL system, which declines weakly towards higher masses, with an overall mean contribution of $59 \pm 13$ per cent. The mass fraction this component makes of the ICL depends heavily on mass, being the largest component of the system at lower masses and declining sharply towards higher masses, with an overall mean contribution of $44 \pm 18$ per cent. 
    \item The in situ component is the second largest component of the BCG+ICL system, which declines weakly towards higher masses and has an overall mean contribution of $28 \pm 9$ per cent. The mass fraction this component makes of the ICL shows the weakest dependence on mass, declining slightly for higher masses, with an overall mean contribution of $15 \pm 6$ per cent. 
    \item The stripped galaxies component is the smallest component of the BCG+ICL system and increases sharply towards higher masses, with a mean contribution of $12 \pm 10$ per cent. The mass fraction this component makes up of the ICL depends heavily on mass, being the largest component of the system at higher masses and declining sharply towards lower masses. The overall mean contribution to the ICL is $41 \pm 20$ per cent.
    \item Clusters with a large surviving fraction in the ICL are more likely to be disturbed and have lower concentrations.
    \item Clusters with a large completed mergers fraction in the ICL are less likely to be disturbed and have higher concentrations
    \item Clusters with a large in situ fraction in the ICL are unlikely to be disturbed and are more highly concentrated. 

\end{enumerate}

\subsection{ICL and BCG properties}
\begin{enumerate}
    \item There is a significant offset in metallicity between the ICL and the BCG, with the ICL being considerably more metal-poor.
    \item The stripped component is significantly more metal-poor than the completed mergers and in situ components
    \item The metallicity gradient across the BCG+ICL is universally negative, \hl{with gradients increasing towards lower masses. The scatter in the relation also increases towards lower masses.}
    \item There is a significant offset in colour between the ICL and the BCG, with the ICL being considerably bluer.
    \item The stripped component is significantly bluer than the completed mergers and in situ components, although it has a positive slope, becoming redder in line with the cluster mass.
    \item The colour gradient across the BCG+ICL is universally negative and shows no particular trend with mass, although the scatter increases considerably towards lower masses
    \item There is no significant difference in age between the ICL and BCG.
\end{enumerate}

\subsection{ICL and BCG progenitors}
\begin{enumerate}
    \item The mean mass of ICL progenitors tracks slightly behind the mean mass of BCG progenitors. The mean mass for both increases with cluster mass. At a given mean mass for the BCG progenitors, higher-mass clusters have higher-mass ICL progenitor galaxies.
    \item The ICL and BCG share a high fraction of progenitor galaxies. Overall, 93 per cent of galaxies that contributed to the growth of the ICL also contributed to the BCG and 91 per cent of galaxies that contributed to the growth of the BCG also contributed to the ICL.
    \item Clusters with shared accretion histories are more likely to be relaxed and more highly concentrated.
    \item Material from high-mass progenitor galaxies is generally more centrally concentrated than material from low-mass progenitor galaxies.
    \item On average, the contribution of a single progenitor to the ICL is $27 \pm 12$ per cent.
    \item The same \hl{most significant
progenitor (i.e. the progenitor that contributed the most
mass) is shared by 36 per
cent of BCG+ICL systems.} 
    \item Clusters dominated by low-mass galaxies in the ICL are likelier to be in an intermediate state and have high concentrations and less recent major mergers.
    \item Low-mass galaxies contributing to the ICL indicate that the ICL is more metal-poor and has a sharper metallicity slope.
    \item Clusters with a high proportion of material from surviving galaxies have higher mean masses of progenitor galaxies and have a lower likelihood of sharing significant progenitors.
    \item Clusters with a high proportion of material from completed mergers were more likely to share significant progenitors and more likely to have a high contribution from a single galaxy.
    \item Clusters with a higher in situ fraction were likelier to share common progenitors and have low-mass progenitor galaxies. 
\end{enumerate}

In this work, we find that the ICL is primarily composed of massive \hl{\mbox{(M\textsubscript{*}~>~1~$\times$~10\textsuperscript{10}~\(\textup{M}_\odot\))}} galaxies experiencing tidal stripping, in contrast to the BCG, which is primarily formed by major mergers and in situ star formation. The ICL and the BCG generally share an accretion history, but in some cases, the accretion histories are distinct. Coupled with the more sizeable in situ component in the BCG, this indicates that BCG properties are sometimes a poor indicator of ICL formation. The ICL mass fraction is correlated with cluster mass, concentration, and time since the last major merger, but with significant amounts of scatter. \hl{Our results show that the formation of the ICL in clusters varies significantly from cluster to cluster and depends strongly on their specific accretion histories. Finally, we recall that the properties of the ICL can be highly sensitive to different methods of defining the separation between the ICL and the BCG. Therefore, special care must be taken when comparing the results of different studies.}

\begin{acknowledgements}
\hl{We thank the anonymous referee for their detailed and constructive comments, which have helped us to improve the paper.}

RJM and FAG acknowledge ANID Fondo 2021 GEMINI ASTRO21-0061. The authors gratefully acknowledge support from the ANID BASAL project FB210003. FAG and AM acknowledge support from the ANID FONDECYT Regular grants 1251493 and 1251882, respectively, as well as funding from the HORIZON-MSCA-2021-SE-01 Research and Innovation Programme under the Marie Sklodowska-Curie grant agreement number 101086388. 
\end{acknowledgements}

\bibliographystyle{aa}
\bibliography{biblio} 

\end{document}